\documentclass[arxiv,usenatbib]{iupartex}
\usepackage{newtxtext,newtxmath}
\usepackage[T1]{fontenc}
\usepackage{ae,aecompl}

\usepackage{multicol}   
\usepackage{bm}		    
\usepackage{pdflscape}	

\title[Mass-Luminosity Relations and Beyond]{Fundamentals of Stars: Critical Looks at Mass-Luminosity Relations and Beyond}

\author[Eker et al.]{%
Z.~Eker$^{1\cc}$\orcid{0000-0003-1883-6255},
F.~Soydugan$^{2,3}$\orcid{0000-0002-5141-7645}
and
S.~Bilir$^{4}$\orcid{0000-0003-3510-1509}
\affsep \\
$^1$Akdeniz University, Faculty of Sciences, Department of Space Sciences and Technologies, 07058, Antalya, Türkiye\\
$^2$\c{C}anakkale Onsekiz Mart University, Faculty of  Sciences, Department of Physics, 17100, \c{C}anakkale, T\"{u}rkiye\\
$^3$\c{C}anakkale Onsekiz Mart University, Astrophysics Research Center and Ulup{\i}nar Observatory, 17100, \c{C}anakkale, Türkiye\\
$^4$ Istanbul University, Faculty of Science, Department of Astronomy and Space Sciences, 34119, Beyazıt, Istanbul, Türkiye\\
}

\corres{Z. Eker}{eker@akdeniz.edu.tr}

\pubyear{2023}

\doiheader{XXXXXXX/PAR.20XX.00000}
\date{
	\pSubmit{XX.XX.XXXX} 
	\pRevReq{XX.XX.XXXX}
	\pLastRevRec{XX.XX.XXXX}
	\pAccept{XX.XX.XXXX}
	\pPubOnl{XX.XX.XXXX}
}
\volume{0}
\volnumber{1}

\begin{document}
\label{firstpage}
\pagerange{\pageref*{firstpage}--\pageref*{lastpage}}
\maketitle

\begin{abstract}
Developments on various relations among stellar variables such as the main sequence empirical mass-luminosity (MLR), mass-radius (MRR) and mass-effective temperature (MTR) relations were reviewed. Conceptual changes on their understanding and usages were discussed. After its discovery, MLR was treated as one of the fundamental secrets of the cosmos, at least until questioned by the end of the 20th century. Differences between fundamental laws and statistical relations were used to understand long-term developments of MLR, together with improvements observed in MRR and MTR. Developments show a break point, initiated by \citet{Andersen1991}, in the line of progress. Before the break when reliable data were limited, MLR and MRR were calibrated using $M$, $L$, and $R$ of binary components of all kinds visual, spectroscopic, and eclipsing for two purposes: {\it i}) obtaining mean mass ($\langle M\rangle$), mean luminosity ($\langle L\rangle$), and mean radius ($\langle R\rangle$), {\it ii}) to estimate $M$ and $R$ of single stars. By the time of the break, the number of solutions from detached double-lined eclipsing binaries (DDEB) giving accurate $M$ and $R$ within a few percent levels are increased. Parameters from very close, semi-detached, and contact binaries were excluded for the sake of refinement, however, MLR and MRR diagrams were found insufficient to derive MLR and MRR functions because the dispersions are not only due to random observational errors but also due to chemical composition and age differences. Then, the new trend was adopted by replacing classical MLR and MRR with empirical $M$ and $R$ predicting relations.Thus, the purpose one was suppressed also because the new trend found a fruitful application in determining $M$ and $R$ of exoplanet hosting single stars. Corrections on misnames and devising new classical MLR, MRR, and MTR, giving mean values ($\langle M\rangle$, $\langle L\rangle$, $\langle R\rangle$) are encouraged since they are still useful and needed by astrophysical models requiring such mean values, not only beneficial to astrophysics, but also beneficial to galactic, extragalactic search, even cosmological models.
\end{abstract}

\begin{keywords}
Stars: fundamental parameters, Stars: luminosity function, mass function, Galaxies: luminosity function, mass function, Cosmology: miscellaneous
\end{keywords}



\section{Introduction}
\label{sec:Introduction}
Accuracy and precision of observational parameters of stars are crucial not only for improving stellar structure and evolution theories but also for fundamental astrophysics as well as Galactic and extragalactic studies, and ultimately even for the cosmological models because stars and galaxies are primary building blocks of the universe. Fundamental and statistical relations are essential for understanding physical events occurring in various parts of the universe. The fundamental relations are the relations like Stefan-Boltzmann law ($L=4\pi R^{2}\sigma T_{\rm eff}^{4}$) which are characterised by at least two properties; {\it i}) how various properties of stars are related, e.g. how the luminosity of a star is related to its radius and effective temperature; {\it ii}) validity is not limited by certain conditions, that is, it is applicable to all stars and radiating surfaces as long as the source of radiation is thermal. The statistical relations, on the other hand, may not work in all possible cases; appropriate statistical conditions are demanded.

A good example is the kinetic temperature. A temperature is a physical quantity that is measured in the macro world but does not exist in the micro world. This is because temperature is a statistically defined quantity implying an average kinetic energy per particle in a substance which could be a solid, liquid, or gas. The definition of kinetic temperature tells us that a single particle or insignificant number of particles cannot be associated with any temperature. For example, when we talk about air temperature, it indicates the average kinetic energy per air particle, which could be written as:

\begin{equation}
\frac{1}{2}m\langle v^{2}\rangle=\frac{3}{2}kT
\end{equation}
where $m$ and $\langle v^{2}\rangle$ are the mean mass and root-mean-square (rms) speed of particles. They are not physical quantities but just average values. Increasing the speed of the substance itself by moving it faster, e.g. putting it on a very fast airplane, does not change its temperature. Therefore, equation (1) cannot be written for a randomly chosen particle (or a few particles), that is, the right-hand side does not exist except for a group of particles satisfying the implied statistic.

Similarly, the mass–luminosity relation (MLR) in the form ($L\propto M^{\alpha}$), which is defined so far for main-sequence stars, is a statistical relation like equation (1) which indicates how main-sequence luminosities are related to main-sequence masses. As it is not valid for non-main sequence stars, depending upon expected accuracy, it may also not be valid for an individual star. So, it is the relation devised for estimating a typical mass from a typical luminosity, or vice versa.

Consequently, inter-related mass-luminosity (MLR), mass-radius (MRR), and mass-effective temperature relations (MTR) of \citet{Eker2018} all are statistical relations, which should not be treated like fundamental relations.  While, bolometric correction – effective temperature relation (BC-$T_{\rm eff}$) of \citet{Flower1996}; BC-$T_{\rm eff}$ and BC-mass relations of \citet{Eker2020, Eker2021a}; \citet{Bakis2022}; \citet{Eker2023} are all also statistical relations which are not to be confused by fundamental relations, otherwise unexpected results or upsetting errors in the computed quantities becomes unavoidable.

One cannot say ``statistical relations are less valuable than the fundamental relations''. On the contrary, a statistical relation could be more valuable; even more practical, or easier to use, e.g. calculating the total energy of a gas is equal to the mean energy per particle multiplied by the number of particles in the gas. Otherwise, the probability of each particle having a certain kinetic energy, and the number of particles having this energy is required before integrating them over all possible kinetic energies. Similarly, using an MLR is more practical to determine the masses and luminosity of galaxies, which are the key parameters to determine dark matter in galaxies. Otherwise, summing individual masses and luminosities of the stars in a galaxy, which are impractical observationally, are needed to obtain the total mass and the total luminosity of the galaxy in question. 

Moreover, there could be various astrophysical research demanding MLR, MTR, and MTR besides stellar astrophysics. Live examples are such as cometary research \citep{Wysoczanska2020b, Wysoczanska2020a}, Oort clouds \citep{Baxter2018}, heliophysics and planetary habitability \citep{Schrijver2019}, exoplanet investigations \citep{Berger2020, Arora2021, Burt2021, Caballero2022, Dattilo2023}, planetary nebula \citep{Munday2020, Alter2020}, open clusters \citep{Ilin2021, Akbulut2021, Yontan2021, Yontan2023a, Yontan2023b}, dark matter searches \citep{Garani2022, Peled2022} and quasars \citep{Albert2021}, neutron stars \citep{Yuan2022}, black holes \citep{Roy2021a, Roy2021b}, general relativity \citep{Lalremruati2021}, gravitational lensing \citep{Ramesh2022, Pietroni2022} and even search for extra-terrestrial intelligence \citep[SETI,][]{Kerins2021, Kerins2023}, which are all used at least one of the statistical relations MLR, MRR and MTR of \citet{Eker2018}.  

Apparently, differentiation between fundamental and statistical relations is important from an astrophysical point of view. Recognising statistical relations is even more important, otherwise using them as fundamental relations would be misleading. Unfortunately, there are authors such \citet{Malkov2003, Malkov2007}, \citet{Henry2004} and \citet{Gafeira2012} including \citet{Eker2015} did not hesitate to call MLR ``fundamental law'', ``sufficiently fundamental to be applicable to many areas of astronomy'', ``one of the most famous empirical law'' and ``one of the fundamental secrets of the cosmos''. On the contrary, \citet{Andersen1991} and \citet{Torres2010} preferred not to define a MLR and preferred displaying the $\log M-\log L$ diagram without a function (MLR) fitting to the data because the scatter on the diagram is not due to observational random errors but most likely abundance and evolutionary effects. \citet{Andersen1991} claims ``....departures from a unique relation is real''. If there is no unique function to represent the data on the diagram, why bother to define one?     

Therefore, this review article is dedicated to investigating the evolution of the statistical functions MLR, MRR, and MTR starting from their discovery until today and trying to explain why such quarrels occurred and whether it is possible to resolve such quarrels by identifying the nature of the relation. Moreover, the presentation of their conceptual advances and realizing whether they were perceived as fundamental laws or statistical relations will benefit to astronomical community. Developing such a conscious analysis would result in a better understanding of their previous usages as well as future studies.

\section{Overview}
Calibrations of MLR, MRR, and MTR require pre-determined accurate stellar parameters such as mass ($M$), luminosity ($L$), radius ($R$), and effective temperature ($T_{\rm eff}$). The most critical parameter among them is $L$ because it is not an observable quantity, that is, there is no telescope nor a detector to be able to observe the total radiation radiated from a star at all frequencies \citep{Bakis2022, Eker2023}. Fortunately, there are only one direct and two indirect methods to calculate $L$ of a star \citep{Eker2021b}. The first method is a direct one because it uses independently determined observational $R$ and $T_{\rm eff}$ of a star to calculate its $L$ directly from its radiating surface area ($4\pi R^2$) and bolometric flux ($\sigma T_{\rm eff}^4$), which is commonly known as the Stefan-Boltzmann law ($L=4\pi R^2\sigma T_{\rm eff}^4$). The other two methods are indirect because the first one providing $L$ of a star from its $M$ requires a pre-determined classical MLR in the form of $L \propto M^{\alpha}$, while the other supplying $L$ of the star from its apparent brightness and distance requires a pre-determined BC-$T_{\rm eff}$ relation. Unfortunately, the indirect methods are useless without their pre-determined relations, and thus, one has no other choice but to use the Stefan-Boltzmann law to produce earliest sample of $L$ values for the first calibration of MLR, MRR, and MTR.     

Assuming that the sample $L$ values are ready, the next most critical parameter is stellar $M$ to establish not only MLR but also MRR and MTR since sequentially required $R$ and $T_{\rm eff}$ should have been already used in computing $L$ of the sample stars. However, like $L$ of a star, $M$ of a star is also not observable. This fact makes the parameter $M$ even more critical since only the masses of binaries (or multiple systems) could be calculated using Kepler’s Third Law from the observed orbital semi-major axis of the components and the orbital period which could be deduced from the observed orbits of visual binaries (or multiple systems) or from the radial velocity curves of double-lined eclipsing spectroscopic binaries. Stellar spectra do not provide orbital inclinations, thus spectroscopic binaries without eclipses cannot provide component masses unless orbital inclinations are available independently.  

Eclipsing binaries are ideal objects to collect observed radii of stars while the single stars are null, except the ones that are close in distance and large enough where interferometry could be useful. Although binarity adds complications to estimating component effective temperatures, the eclipsing binaries are still advantageous to revealing the most accurate effective temperatures and temperature ratios from the depths of minima if the effective temperature for one of the components is estimated correctly. 

Visual binaries (multiple systems too) and double-lined eclipsing binary systems are the only objects provide $M$ from Kepler’s third law, with $R$ from eclipses. On the other hand, there is almost no other way to obtain accurate masses and radii of single stars except seismic analysis obtained mass and radii data for especially solar-like pulsating stars \citep[e.g.][]{Gaulme2016, Bellinger2019}. Considering some basic information and data knowledge about this research topic, we must now start reviewing the statistical relations starting from the most prominent one: MLR. 

\subsection{Revisiting MLR}
The famous stellar mass-luminosity relation (MLR) was discovered empirically in the middle of the first half of the 20th century by \citet{Hertzsprung1923} and \citet{Russell1923} independently from the masses and absolute brightness of a very limited number of visual binaries. Eclipsing binaries were included later in the statistics. In his MLR, \citet{Eddington1926} was able to use 13 eclipsing binaries together with 29 visual binaries and five cepheids, which was available to him at that time, while \citet{Mclaughlin1927} increased the number of eclipsing systems to 41 in his plots. 

Almost immediately after discovery, \citet[][ page 154]{Eddington1926} fit a curve to the data on a diagram $\log M - M_{\rm Bol}$ ($\log$ mass versus absolute bolometric magnitude) and call it ``the theoretical mass-luminosity law''. Examining masses and bolometric magnitudes of 18 visual binaries, \citet{Gabovits1938} agreed \citet{Eddington1926} after twelve years by re-examining the mass-$M_{\rm Bol}$ data again and declared ``We conclude that, as revealed by our selected first-class data, the stars (chiefly of the main-sequence) probably follow a strict mass-luminosity law''. On the other hand, having larger data sets including visual binaries, spectroscopic binaries, Hyades and Trumpler stars and white dwarfs, \citet{Kuiper1938} was rather suspicious to accept the mass-luminosity relation as a law because he commented ``It is doubtful whether this mean relation has any physical significance'' after discussing them on a $\log M- M_{\rm Bol}$ diagram.

MLR has been updated and revised many times, until a major break occurred at the very beginning of the last decade of the 20th century on the issue of whether it is a statistical relation or a fundamental law. Looking at the developments before this break would be useful to understand it better. \citet{Petrie1950a, Petrie1950b} used 93 spectroscopic binary systems, \citet{Strand1954} 
 studied 23 visual binaries, \citet{Eggen1956} investigated 34 visual binaries, \citet{McCluskey1972} considered 40 visual binaries and 35 eclipsing systems, \citet{Cester1983} gathered 45 visual and 40 spectroscopic binaries, \citet{Griffiths1988} analyzed 72 detached main-sequence binaries, 25 detached OB, six resolved binaries and 23 visual binaries when revising the MLR relations. \citet{Demircan1991} preferred to study masses and luminosities of 70 eclipsing binaries (140 stars) only, including the main sequence components of detached and semi-detached binaries as well as the components of OB-Type contact and near contact binaries. \citet{Karetnikov1991} used 303 eclipsing systems of different types. At this point, we must keep in mind that the observational data was very limited, therefore, authors combined different kinds of binaries whether they are eclipsing or not without differentiating between detached, semi-detached, and contact systems. This was done to increase the statistical reliability of MLR's calibrated. 

Those early generation relations, including the very earliest ones, were demonstrated as mostly mass-absolute magnitude diagrams, some are with the best fitting function, and some are without a fitting curve. Considering the classical form  $L\propto M^{\alpha}$ of MLR, first \citet{Eggen1956} intended to define the power of mass ($\alpha$) so he expressed $L= \mu^{3.1}$, where $\mu$ is the total mass of a double-star system defined as $\mu= a^3/P^2\pi ^3$ from the Kepler’s Harmonic Law. Then, \citet{McCluskey1972} preferred to use a relation in the form $M\propto L^{\beta}$, where $M$ and $L$ are the masses and luminosities of the components while $\alpha$ and $\beta$ are the constants to be determined by the data on various mass-absolute bolometric magnitude diagrams. Accordingly, \citet{Cester1983}; \citet{Griffiths1988} and \citet{Demircan1991} preferred to study MLR on a mass-luminosity diagram for defining unknown constants on the classical form ($L\propto M^{\alpha}$) of MLR either by fitting a curve to all data or dividing the mass range into two regions as the low and the high mass stars, or three regions as the high, intermediate or Solar, and low mass, in order to determine the inclinations (power of $M$) and it is zero point constants of the linear MLRs on the $\log M - \log L$ diagram. 

\subsubsection{Is MLR a Fundamental Relation?}
The major brake on the concept and splitting practices on the purpose of calibrating a MLR is triggered by \citet{Andersen1991} who collected 45 detached double-lined eclipsing binary (DDEB) systems (90 stars) having both masses and radii accurate within 2\%, which were the most accurate stellar data of the time. Preferring only DDEB stars was not intended to reject the reliability of the other sources for determining masses and radii; only because $M$ and $R$ from DDEB were considered as the most reliable. 

\citet{Andersen1991} rejected calibrating any form of MLR function to fit data on the $\log M - \log L$ diagram. The diagram displayed by him is without a fitting curve (MLR) because the scatter from the curve is certainly not only due to random observational errors but also due to abundance and evolutionary effects. He exclaimed: ``At first glance, the mass-luminosity diagram shows a tight, well-defined mean relation. Closer inspection, taking the individual uncertainties into account, reveals that the departures from a unique relation are real''.  Then, if there is no unique curve expressible by a function to represent data, why bother to draw one?

In very early times, especially after its discovery, MLR was claimed to be one of the most prominent empirical laws of nature by \citet{Eddington1926} and \citet{Gabovits1938} and commonly used by researchers either reckoning $L$ of a star from its $M$ or estimating $M$ of a star from its $L$ at least until the middle of the 20th century, and perhaps until \citet{Andersen1991} objecting it. This practise was extremely useful for single stars because there was no other observational way to access their masses directly, but the absolute bolometric magnitudes of the ones with known parallaxes were rather easy to get by using a proper bolometric correction (BC), and then estimate their masses from a pre-determined bolometric magnitude-mass relation. The method was even helpful to double-check trigonometric parallaxes of single stars and even eclipsing binaries with light curve solutions giving $R$ and $T_{\rm eff}$ of the components, thus $L$ values are compatible by the Stefan-Boltzmann law. Those very early times were a period in astrophysics when nuclear reactions were not fully established and understood correctly. The discovery of nuclear reactions in the cores of stars are attributed to Sir Arthur Eddington \citep{Bahcall2000} a few years before the discovery of MLR by \citet{Hertzsprung1923} and \citet{Russell1923}.  

Astronomers waited until 1932 for the discovery of neutrons \citep{Chadwick1933} to study and fully understand hydrogen fusion as a main source of stellar energy. Only after the CNO cycle was established by Hans Bethe and Von Weizsacher \citep{Clayton1968}, and only a year later the p-p chain reactions were suggested by \citet{Bethe1939}, were the solutions of stellar structure equations with the nuclear energy which placed our theoretical understanding of the evolution of stars with evolutionary tracks on a solid ground \citep{Clayton1968}. Empirically discovered MLR was confirmed later theoretically in a sense that mass ($M$) is the prime parameter that determines the internal structure, size ($R$), and luminosity ($L$) of a star not only for the time span of the main sequence but also throughout the star’s lifetime until its death, where its initial chemical composition can cause little variations. So, the scatter on a $\log M - \log L$ diagram for field main-sequence stars is not only due to observational errors but also due to various ages and chemical compositions. Additionally, at least until the middle of the 20th century, maybe until about the middle of its second half, but certainly not until \citet{Andersen1991}, the observational accuracy was not high enough to differentiate theoretical (true) $L$ and $M$ of stars which are marked on an H-R diagram to form evolutionary tracks. As long as observational accuracies of $L$ and $M$ are very low that the error bars covering the thickness of the main sequence, then astronomers did not suspect inconsistency between the predicted and the observed quantities using a MLR. It was normal for them to be satisfied as in the case of fundamental law. 

By the time of \citet{Andersen1991}, who collected the most accurate $M$ and $L$ data from DDEB, there were sufficiently accurate masses and luminosities thus one can deduce the scatter on the $\log M - \log L$ diagram are not only due to observational errors of $M$ and $L$ but also due to abundance and evolutionary effects. \citet{Andersen1991} felt something was not right, that is, there must be a problem in treating classical MLR in the form $L \propto M^{\alpha}$ as a fundamental relation. This was because, Andersen thought, such a fundamental relation must not only fit the data uniquely but also must contain the other parameters involving chemical composition and evolution of the stars on the diagram.  

Being influenced by the common usage, a kind of paradigm not easy to be free off, \citet{Andersen1991} was expecting, like the others before him, to obtain the mass of a single star from its luminosity within acceptable accuracy [ideally $\pm$5\% see \citet{Andersen1991} page 93]. However, with a mean MLR in the form $L \propto M^{\alpha}$ for the main-sequence stars, this was not possible anymore due to improved observational techniques with higher accuracy in both $L$ and $M$ values available to him. Nevertheless, he succeeded in this aim about two decades later by replacing $\log L$ with a function including observational parameters $T_{\rm eff}$, surface gravity ($\log g$), and relative iron abundance [Fe/H] to incorporate evolution (age) and chemical composition in a new review paper lead by \citet{Torres2010} as
\begin{equation}
    \log M = a_1 + a_2X + a_3X^2 + a_4X^3 + a_5(\log g)^2 + a_6(\log g)^3 + a_7{\rm [Fe/H]}, 
\end{equation}
where $X\equiv \log T_{\rm eff} - 4.1$, which is the parameter contains the most effective variable, $T_{\rm eff}$, directly related to the luminosity with its fourth power in the Stefan-Boltzmann law, and $a$’s which are the calibrated coefficients $a_1 = 1.5689\pm0.058$, $a_2=1.3787\pm0.029$, $a_3=0.4243\pm 0.029$, $a_4=1.139\pm0.240$, $a_5=-0.1425\pm 0.011$, $a_6=0.01969\pm0.0019$, and $a_7=0.1010\pm 0.058$. This is a function providing 6.4\% accuracy for main-sequence and evolved stars above $0.6M_{\odot}$. It is obvious that such a function cannot be drawn on a $\log M-\log L$ diagrams but provides stellar $M$ as the classical MLR but much more accurate not only for main-sequence stars but also giants and sub-giants thus, both \citet{Andersen1991} and \citet{Torres2010} presented their $\log M-\log L$ diagrams without a fitting function. 

Incorrect diagnoses are natural to be continued by incorrect treatments. First of all, classical MLR in the form of $L \propto M^{\alpha}$ is a statistical relation devised for estimating a typical mass from a typical luminosity, or vice versa, not for estimating $M$ or $R$ of a single star, even if it was once used to estimate $M$ of a star from its $L$. Therefore, It is not right to diagnose the classical MLR as one of the fundamental relations to calculate the mass of a star from its luminosity. It is not right to look for a relation giving the mass of a star from other stellar observational parameters for replacing MLR. It is not right to call this relation MLR if it provides $M$ of a star from the other parameters even if $L$ is included in the right-hand side of the equal sign except if the right-hand side contains only $L$ as a variable. It is not right to claim that there is no uniquely fitting function to the data on $\log M - \log L$ the diagram despite a tight, mean relation between $M$ and $L$ is obvious as if the least squares method would fail to produce one.       

Nevertheless, \citet{Andersen1991}'s exclamation had a noticeable consequence in stellar astrophysics. It appears as if a main cause of deviation on understanding, definition, and usage of newly defined MLR functions from the main path which was continued by \citet{Ibanoglu2006} and \citet{Eker2015, Eker2018} where statistical relation between masses and luminosities of main-sequence stars was kept in the form $L \propto M^{\alpha}$.  

\subsubsection{Mass Predicting Relations of the Deviated Path}
\citet{Andersen1991}'s objection was very effective in the literature that some authors \citep{Malkov2003, Malkov2007, Torres2010} also presented their empirical mass-luminosity diagrams without a curve fitting to the data. \citet{Gafeira2012} just gave mass-luminosity relations for main sequence FGK stars without displaying them on a mass-luminosity diagram. \citet{Fernandes2021}, who compared their results to \citet{Torres2010}, could be considered an improved version of \citet{Gafeira2012}, thus also did not display mass-luminosity relations on the mass-luminosity diagram.  

Various kinds of mass-predicting empirical relations were calibrated and most of them were erroneously called MLR. \citet{Gorda1998}, who preferred the form $M_{\rm Bol}=a+b\log M$, where $M_{\rm Bol}$ is the absolute bolometric magnitude and $M$ is the mass. \citet{Henry1993}, who preferred $\log M = aM_{\xi}+b$ for infrared colors, where $M_{\xi}$ indicates absolute magnitudes at the $J$, $H$, and $K$ bands, and $\log M = aM_{\rm V}^2+bM_{\rm V}+c$ for the $V$ band to express various MLR with unknown coefficients $a$, $b$, and $c$ to be determined by the data on various diagrams. The former relation in the form $M_{\rm Bol}=a+b\log M$ could be justified to be named mass-luminosity relation since $M_{\rm Bol}$ of a star is directly related to its luminosity, but a relation in the form $\log M = aM_{\xi}+b$, is definitely not a mass-luminosity relation. Such relations should better be called mass-absolute brightness relations to avoid confusion. 

Style of expressing $V$-band mass-absolute brightness relation as second-degree polynomials covering masses $0.6~M_{\odot}$ to 22.89 $M_{\odot}$ is continued by \citet{Xia2010} who calibrated it within two regions one for $M_{\rm V} < 1.05$ mag ($2.31<M/M_{\odot}<22.89 $) and other for $M_{\rm V}>1.05$ mag ($0.60< M/M_{\odot}<2.31$) using the dynamical masses and $V$-band absolute magnitudes of 203 main-sequence stars, but still the relation is called MLR. The accuracy of predicted masses is estimated to be within 5\%. The style of displaying mass ($M/M_{\odot}$) versus absolute magnitude data, where the absolute magnitudes are in the visual and the $K$-bands is continued by \citet{Benedict2016} for the low end of the main sequence ($M<0.6M_{\odot}$) but in different functional forms predicting masses as accurate as $\pm0.035M_{\odot}$ in the region $M=0.2M_{\odot}$. Similarly, both direct (absolute magnitude for a given mass) and inverse (mass for a given absolute magnitude) relations of \citet{Benedict2016} are better to be called empirical mass-brightness relations rather than mass-luminosity relations. The empirical data of \citet{Benedict2016} are also compared to the theoretical mass-absolute brightness curves of age 1 Gyr of \citet{Baraffe2015} and \citet{Dotter2016}. $K$-band data was found to fit better than the $V$-band data.      

Mass-Absolute visual magnitude $[M_{\rm V}-(\log m)]$, mass-luminosity $[\log L-(\log m)]$, mass-temperature $[\log T_{\rm eff}-(\log m)]$, and mass-radius [$\log R-(\log m)]$ relations and corresponding inverse relations are calibrated for the intermediate-mass stars in the mass range $1.4< M/M_{\odot}< 12$ by \citet{Malkov2007} in forms of polynomials with empirically determined coefficients from the fundamental parameters of stars collected from DDEB and visual binaries. Similar empirical relations were studied later in a wider perspective by \citet{Moya2018} using a set of observational parameters $M$, $R$, $T_{\rm eff}$, $g$ (surface gravity), $\rho$ (mass density), and [Fe/H] of 934 stars of eclipsing binaries and single stars observed by asteroseismology and interferometry where the two-thirds of the stars are on the main sequence. A total of 576 linear combinations of $T_{\rm eff}$, $L$, $g$, $\rho$, and [Fe/H] (and logarithms) were used as independent variables to estimate $M$ and $R$, but \citet{Moya2018} presented only 38 of them with regression statistics adj-$R^2$ higher than 0.85. Accuracy better than 10\% was achieved in almost all cases of 38 equations. The term ``empirical relations'' used by \citet{Moya2018}, and the names of relations given by \citet{Malkov2007} are proper and consistent in comparison to the names used for the mass predicting relations in the previous paragraph.

Using $\log L/L_{\odot}$, [Fe/H] and star age/age$_{\odot}$ as free variables, \citet{Gafeira2012} suggested three different equations so that one can choose one according to the availability of data. The simplest is the first one which is a third-degree polynomial without a constant term, having $X$, where $X\equiv \log L/L_{\odot}$. The second equation adds [Fe/H] as the new variable in another third-degree polynomial into the equation in addition to the first. The third equation combines three third-degree polynomials, where the third uses the relative age (age/age$_{\odot}$) of the star as a variable. With all parameters, the third relation looks like equation (2) above by \citet{Torres2010}. That is, the part $a_1 + a_2X + a_3X^2 + a_4X^3$ is replaced by $0.0219(\pm 0.023)\log L/L_{\odot} + 0.063(\pm 0.060)(\log L/L_{\odot})^2-0.119(\pm 0.112)(\log L/L_{\odot})^3$ while the part $a_7$[Fe/H] is replaced by $+0.079(\pm 0.031){\rm [Fe/H]}-0.122(\pm 0.119){\rm [Fe/H]}^2-0.145(\pm 0.234){\rm [Fe/H]}^3$ at last the part $+a_5(\log g)^2+a_6(\log g)^3$ is replaced by $+0.144(\pm 0.062) ({\rm age/age_{\odot}})-0.224(\pm 0.104)({\rm age/age_{\odot}})^2-0.076(\pm 0.045)({\rm age/age_{\odot}})^3$. Adding age and metallicity improved the mass estimation (15\% to 5\%) for FGK stars \citep{Gafeira2012}. However, only the first equation could be called MLR, but not the other two, which are better to be called mass-luminosity-metallicity relation and mass-luminosity-metallicity-age equation respectively. Nevertheless, \citet{Fernandes2021} corrected this by changing the name ``the mass and radius-luminosity-metallicity-age relations'' in the new version which was calibrated by 56 stars with metallicity and mass in the ranges ${\rm -0.34<[Fe/H] (dex) <0.27}$ and $0.66<M/M_{\odot}<1.8$ from the DEBCat catalogue\footnote{https://www.astro.keele.ac.uk/jkt/debcat/ }. Estimated accuracy in the new version is 3.5\% and 5.9\% in predicting $M$ and $R$ of single stars, respectively. \citet{Serenelli2021} commented ``\citet{Gafeira2012} provided three relations for the stellar mass, but only two of them can be easily applied''. 

\citet{Serenelli2021} studied mass determination methods from a wider perspective including all existing techniques older and newer; involving spectroscopy and/or photometry; theoretical or observational and at last summarized them in a figure \citep[Fig. 16 of][]{Serenelli2021} graphically where one can see their applicable ranges of stellar mass as well as the accuracy and precision of the predicted masses. Despite deceiving examples as summarized in the three paragraphs above, \citet{Moya2018} were careful not to use the word ``MLR'' for naming their ``empirical relations'', which are established to estimate $M$ and $R$ only. While \citet{Fernandes2021} constantly called their simiar functions MLR and claimed ``for single nearby Solar-type stars, the luminosity can be obtained observationally, but not the mass'' opposing the fact that the luminosity of a star is actually not an observable parameter because there is no telescope/detector to observe at all wavelengths. Calling these $M$ or $R$ predicting relations ML/MR would be a main source of confusion among readers. For example, ``empirical ML/MR relations'' marked in Figure 16 of \citet{Serenelli2021} at an accuracy/precision in between 10\% and 15\% for stars $M<2.5M_{\odot}$, where in the caption ``ML/MR for mass-luminosity and mass-radius relations'' written clearly, could be indeed confusing to a careful reader.

\begin{figure}
\centering
\includegraphics[scale=0.50, angle=0]{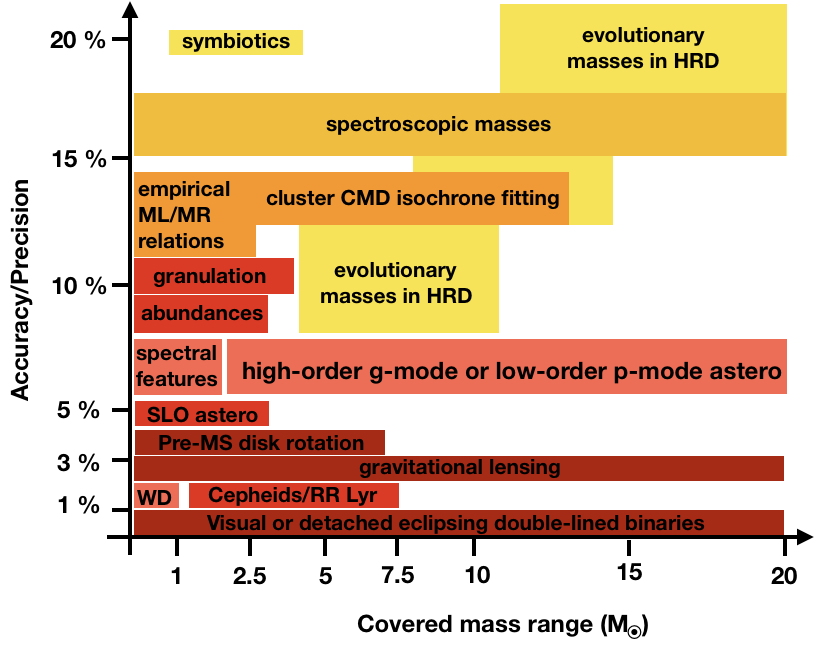}
\caption{~Mass ladder summarizing the capacity of various methods to obtain stellar masses \citep[credit to][]{Serenelli2021}.} 
\label{fig:Mass_Accuacy}
\end {figure}

This is because: Is it possible \citet{Serenelli2021} to use ML/MR to indicate both of the empirical relations of \citet{Moya2018} and the classical mass-luminosity/mass-radius relations (MLR/MRR) of \citet{Eker2018}? According to Figure~\ref{fig:Mass_Accuacy} \citep[Figure 16 of][]{Serenelli2021} the answer would be ``yes'' because the relations of \citet{Moya2018} are not marked as ``empirical relations'' on the figure, and it is not possible to forget them because they cover one of the very important sections of the review (Section 4.4) where the $M$ predicting relations of other authors are summarized and compared to their counterparts in \citet{Moya2018} (see Table~\ref{Tab:comp_emp_rel}). However, \citet{Serenelli2021} declared ``All the relations except two (those with the largest number of dimensions) have precisions better than 5\%''. If the answer is "yes", then, another problem arises: ``If ML/MR in Figure~\ref{fig:Mass_Accuacy} implies both classical MLR/MRR and the new M/R predicting relations, why the precision for the ML/MR is marked to be from 10\% to 15\% while M/R predicting relations are said to have precisions better than 5\%?''. 

It is possible ML/MR in Figure~\ref{fig:Mass_Accuacy} were actually marked from 3\% to 15\%, but this region on the figure is shadowed by other various methods of predicting stellar mass. Then another conceptual problem comes out:  Only similar quantities could be compared. it is not right to compare the accuracy/precision of the empirical relations of \citet{Moya2018}, which are solely devised for predicting masses of single stars, to the accuracy/precision of the classical MLR, which is primarily devised to establish as a statistical relation between typical masses and luminosities of main-sequence stars. It is inconsistent and meaningless to compare E18 to M18 in Table 9 of \citet{Serenelli2021} (see Table~\ref{table:Serenelli_table}). Similarly, it is scientifically inconsistent if empirical ML/MR in Figure~\ref{fig:Mass_Accuacy} indicates classical MLR/MRR as well, rather than the empirical relations of \citet{Moya2018}. This is because \citet{Eker2018} declared ``(classical) MRR and MTR functions, as well as the MLR functions, are needed by the astronomical community for practical purposes. Those include the need to be able to estimate a typical luminosity, radius, and $T_{\rm eff}$ for main-sequence stars of a given mass''. Which is definitely different than the purpose of predicting $M$ or $R$ of single stars all over the H-R diagram? 

After having $M$ and $R$ predicting relations, further calculations are needed for a meaningful comparison. First calculate $M$, $R$, and then $L$ (if not used as input variable) for main-sequence stars only. Then, plot $\log M - \log L$ and $\log M - \log R$ diagrams, and fit preferred functions (ML or MR) using the least squares method. At last ML/MR becomes comparable to MLR/MRR of \citet{Eker2018}. Why bother to do that, if there are many reliable direct methods of obtaining $M$, $R$, and $L$ of main-sequence stars within accuracy 1\% and a few \% respectfully from radial velocity and light curves of DDEB stars which are marked in the bottom of Figure~\ref{fig:Mass_Accuacy} covering the full ranges of stellar masses. Thus, ML/MR is incompatible with classical MLR/MRR (Figure~\ref{fig:Mass_Accuacy}).  

\begin{table*}
\tabcolsep=6pt
\centering
\small
\caption{~Mass predicting empirical relations of other authors are compared to counterparts in \citet{Moya2018} \citep[credit to][]{Serenelli2021}.
\label{Tab:comp_emp_rel}}
	\begin{tabular}{clcclccc}
		\hline \hline
		Ref. & \multicolumn{1}{c}{Relation}    &  Acc/Prec & Ref. & \multicolumn{1}{c}{Corresponding relation}  &   Acc/Prec\\
		\hline
T10 &  $M=f(X,X^2,X^3,{\rm log}^2g, {\rm log}^3g,{\rm [Fe/H]})$&  7.4/52.9 & M18 & $M= f(T_{\rm eff}, {\rm log}g, {\rm [Fe/H]})$ & 7.5/3.4\\
G12 & $M=f({\rm log}L, {\rm log}^2L, {\rm log}^3L$) & 14.0/0.6 & M18 & ${\rm log}M = f({\rm log}L)$ & 10.1/0.1\\
G12 & $M= f({\rm log}L, {\rm log}^2L, {\rm log}^3L, {\rm  [Fe/H]}, {\rm [Fe/H]}^2, {\rm [Fe/H]}^3)$ &  8.9/0.8 &  M18 & ${\rm log}M = f({\rm log}L, {\rm [Fe/H]})$ & 9.9/0.9\\
M07 & $M=f({\rm log}L,{\rm log}^2L)$ & 11.2/--- & M18 & ${\rm log}M = f({\rm log}L)$ & 10.08/0.13\\
E18 & ${\rm log}L=f({\rm log}M)$ & 33.3/6.9 & M18 & ${\rm log}L=f({\rm log}M)$ & 31.9/0.6\\
		\hline
	\end{tabular}
 \\
\begin{minipage}{17cm}
References: T10 \citep{Torres2010}, G12 \citep{Gafeira2012}, M07 \citep{Malkov2007}, E18 \citep{Eker2018}, M18 \citep{Moya2018}.
\label{table:Serenelli_table}
\end{minipage}
\end{table*}

\subsubsection{Recognition of statistical MLRs}
The classical form of MLR ($L\propto M^{\alpha}$) appreciated by \citet{Cester1983}, \citet{Griffiths1988}, \citet{Karetnikov1991} and \citet{Demircan1991} before the break initiated by \citet{Andersen1991}. This classical form continued by \citet{Ibanoglu2006} when comparing mass-luminosity relations for detached and semidetached Algols despite \citet{Andersen1991}'s exclamation. The most recent examples are by \citet{Eker2015, Eker2018}. The classical or any other form reducible to the classical ($L\propto M^{\alpha}$) has the advantage of being easy to interpret with a value given to the power of $M$, which is known to change by energy generation rate per star mass at cores of stars. Thus, the derivative of MLR function ($dL/dM$), the inclination of a line, on the $\log M - \log L$ diagram is the value of alpha. On the other hand, the real advantage is not only that it works in both directions ($M$ from $L$, or $L$ from $M$), but also because it permits one to relate typical masses and luminosities of main-sequence stars in general.

Obviously, the statistically determined relation between $M$ and $L$ for the main-sequence stars is not true for a single star. This is because $L$, $R$, and $T_{\rm eff}$ of a star change by time, while $M$ of the star stays constant (an evolutionary effect) since mass loss of main-sequence stars (especially for the ones cooler than B spectral types) are too small thus mass loss, is usually ignored \citep{Daszynska-Daszkiewicz2019, Bressan2012}. Moreover, there is a metallicity effect that also changes $L$, $R$, and $T_{\rm eff}$ slightly, therefore, stellar structure and evolution models use chemical composition and $M$ as the two basic free parameters for computing $L$, $R$, and $T_{\rm eff}$, which stands for an output of an internal structure model to be confirmed externally. Finally, it can be concluded that any kind of relation between $M$ and $L$ of a single star cannot easily be deduced from evolutionary tracks or from isochrones. However, the distribution of available stellar luminosities on H-R diagrams or on $\log M- \log L$ diagrams clearly shows that there must exist, at least a statistical, one-to-one relation between a typical mass and typical luminosity of main-sequence stars. But, on the other hand, if $L$ and $M$ were totally independent, identification of main-sequence stars on the H-R diagram would not be possible. In fact, the main-sequence stars were first recognized just according to their extraordinary positional appearance even on very primitive H-R diagrams. Because of their distinct positions, they were named main-sequence stars which is still actively used. After all, it is obvious to everyone now, that as soon as main-sequence luminosities are placed on a $\log M-\log L$ diagram, a one-to-one relation between $M$ and $L$ shows itself clearly with a high-level statistical significance.      

What actually happened so far is that; the statistical relation between stellar $M$ and $L$ for main-sequence stars is so strong and obvious that it was discovered even before the stellar structure and evolution theory was fully established. That was the first reason why it had been evaluated as one of the fundamental secrets of the cosmos like the H-R diagram itself (why are stars not distributed evenly on the surface of the H-R diagram but mostly gathered on the main sequence?). The second reason was that the observational accuracy of those early years of astrophysics was not sufficient to distinguish it as a statistical relation, so it was treated as a fundamental relation even though some serious objections mentioned as above have occurred. Even so, no one has yet clearly declared that MLR is just a statistical relation. 

\citet{Andersen1991} totally rejected the existence of a real relation between $M$ and $L$ of the main-sequence stars because according to him a unique relation (like a Planck law, keeping $T_{\rm eff}$ constant, the deviations from the Plank curve are only due to random errors of observed intensities) does not exist since deviations from a fitting curve (MLR) is not just only due to random observational errors of $L$ and $M$, but also due to chemical composition and age differences. \citet{Eker2015, Eker2018} defended existence of a MLR relation by claiming any sample of data could be expressed by a unique relation because a unique fit of a curve on sample data is guaranteed by the least squares method.

However, this defence occurred intuitively because the distinction between MLR to be a fundamental or a statistical relation was not yet fully founded. Even \citet{Eker2015} said ``One of the fundamental secrets of the cosmos, the famous stellar mass–luminosity relation (MLR), was discovered empirically ...''. Only later, \citet{Eker2018} declared ``The main-sequence MLR is one of the fundamentally confirmed and universally recognized astronomical relations''.

How to solve non-uniqueness problems attributed to the light curves of spotted stars, which is a kind of problem \citet{Andersen1991} pointed out against MLR, was discussed first by \citet{Eker1999}. The same principles adopted by \citet{Eker2018} for defending the uniqueness of MLR, MRR, and MTR functions from the most accurate masses and luminosities of 509 main-sequence stars as the components of DDEB in the Solar neighbourhood of the Milky Way. There could be three types of non-uniqueness problems. According to \citet{Eker2018}, there is no non-uniqueness problem of type I because the preferred function expressing MLR is a power law ($L\propto M^{\alpha}$). There should not be a non-uniqueness problem of type II also because there exist many methods like the least squares for one to achieve a unique fit and determine its coefficients uniquely. Finally, the problem of type III does not exist either because the parameter space of MLR is so simple that there is always one correlation between $L$ and $M$, that is, there is only one $L$ value for a given $M$ and vice versa in the case of inverse MLR. The most general approach to non-uniqueness problems recently applied to main sequence BC-$T_{\rm eff}$ relations by \citet{Eker2021a} when discussing the chronic zero-point problems of the BC scale, which is important for obtaining accurate $L$ of single stars from apparent magnitudes if distances are known.   

Apparently, there is no non-uniqueness problem associated with classical MLR, but still one may run into problems of obtaining accurate $L$ from a given $M$ or vice versa. Such problems are inevitable since classical MLR is not devised solely for obtaining accurate $L$ from an accurate $M$ or vice versa. The main obstacle for obtaining accurate $M$ of a single star from its $L$ using classical MLR is not because of the absence of a unique function \citep{Andersen1991}, but because of the degeneracy induced by the stellar structure and evolution theory \citep{Eker2018}. This is because, theoretically, there are an infinite number of $L$ values for a main-sequence star of given $M$ depending upon its chemical composition and age. Thus, varying chemical composition and ages are not there to cause non-uniqueness to an existing MLR. On the contrary, the chemical composition and age of a star is there to break inferred degeneracy for obtaining accurate $L$, $M$, and $R$, which appear on classical MLR and MRR functions as only variables. For this, one must conduct further investigation on evolutionary tracks. Knowing the mass and the chemical composition of a star, the correct track will be chosen. By knowing the age, both $L$ and $R$ will be read on the track. If the age is not known, then either $L$ or $R$ must be known to read the age of the star on the track. Stellar structure and evolution theory is even useful to determine the accurate mass of a single star from its accurately determined [Fe/H] and $L$, where the value of $R$ could also be obtained if $T_{\rm eff}$ of the star is known accurately.    

Without distinguishing them from a fundamental relation, \citet{Eker2015} calibrated four linear MLR functions covering mass range $0.38<M/M_{\odot}< 32$, using masses and luminosities of 268 main-sequence stars selected from 514 stars as the components of 257 DDEB from the catalogue of \citet{Eker2014}. Three distinct break points separating the four distinct mass domains on a $\log M - \log (L/M)$ diagram were identified. The mass domains were named as low mass ($0.38<M/M_{\odot} \leq 1.05$), intermediate mass ($1.05 < M/M_{\odot} \leq2.4$), high mass ($2.4<M/M_{\odot} \leq 7$) and very high mass ($7 <M/M_{\odot} \leq 32$). The linear MLRs of the four mass domains were compared to linear and quadratic MLR of the full range $0.38<M/M_{\odot}<32$ and the four-piece linear MLRs were found best to represent the data on a $\log M-\log L$ diagram. The break points separating the mass domains were interpreted as abrupt changes in the power of $M$, most probably due to changing the type of efficient nuclear reaction operating in the cores of main-sequence stars.  

The statistical nature of MLR was sensed but not fully grasped yet by \citet{Eker2018}. The referees were against calibrating new MLRs which are useless in predicting $M$ and $R$ of single stars. The new trend in determining MLRs was to use them to obtain $M$ and $R$ of single stars as done by \citet{Andersen1991}, \citet{Henry2004}, \citet{Malkov2007}, \citet{Torres2010}, and \citet{Gafeira2012}. Therefore, the justifications for re-calibrating them again after only three years were explained by \citet{Eker2018} as: There are two tables in the handbook of astronomers, commonly known as ``Allen’s Astrophysical Quantities'' \citep{Cox2000}. The first of the two tables list calibration of MK spectral types with seven columns spectral type, $M(V)$, $B-V$, $U-B$, $V-R$, $R-I$, $T_{\rm eff}$ and BC (Table 15.7 on page 388). The second table with six columns Sp type, mass, radii, surface gravity, mean mass density and rotational speed (Table 15.8 on page 389) is actually the continuation of the first table. The columns of the two tables, thus, were connected by spectral types. The second table has a notification mark to say ``columns containing uncertain values'', which actually indicates statistically determined typical masses, radii, surface gravity and mean mass for a given spectral type. The rationale of the new paper was obvious that interrelated MLR, MRR, and MTR would be very useful in supplying demanded reliable statistical information to the astronomical community.	

After three years, \citet{Eker2018}, this time, calibrated a six-piece MLR covering the mass range $0.179<M/M_{\odot}< 31$, using 509 main-sequence stars selected from 639 stars as the components of 318 pair (DDEB) and one detached spectroscopic triple. One should not be surprised by the higher increasing rate of the calibrating stars, which appears to be 90\% while the number of newly added DDEB binaries is only 24\%. This is because stars with less accurate $M$ and $R$ up to 15\% were used in the calibrations rather than up to 3\% as in the previous study. Thus, \citet{Eker2018} concluded that it is not always good to eliminate less accurate data for a better study because it may mean loss of information rather than gain. Lowering limiting relative accuracy to 15\% was a real contribution to extending the low mass limit down to $0.179 M_{\odot}$ from $0.38M_{\odot}$ and discovery of two more break points on the $\log M-\log L$ diagram. That is, adding new DDEB stars (24\%) to the list was not effective as much as lowering the limiting accuracy.

\begin{figure}
\centering
\includegraphics[scale=0.50, angle=0]{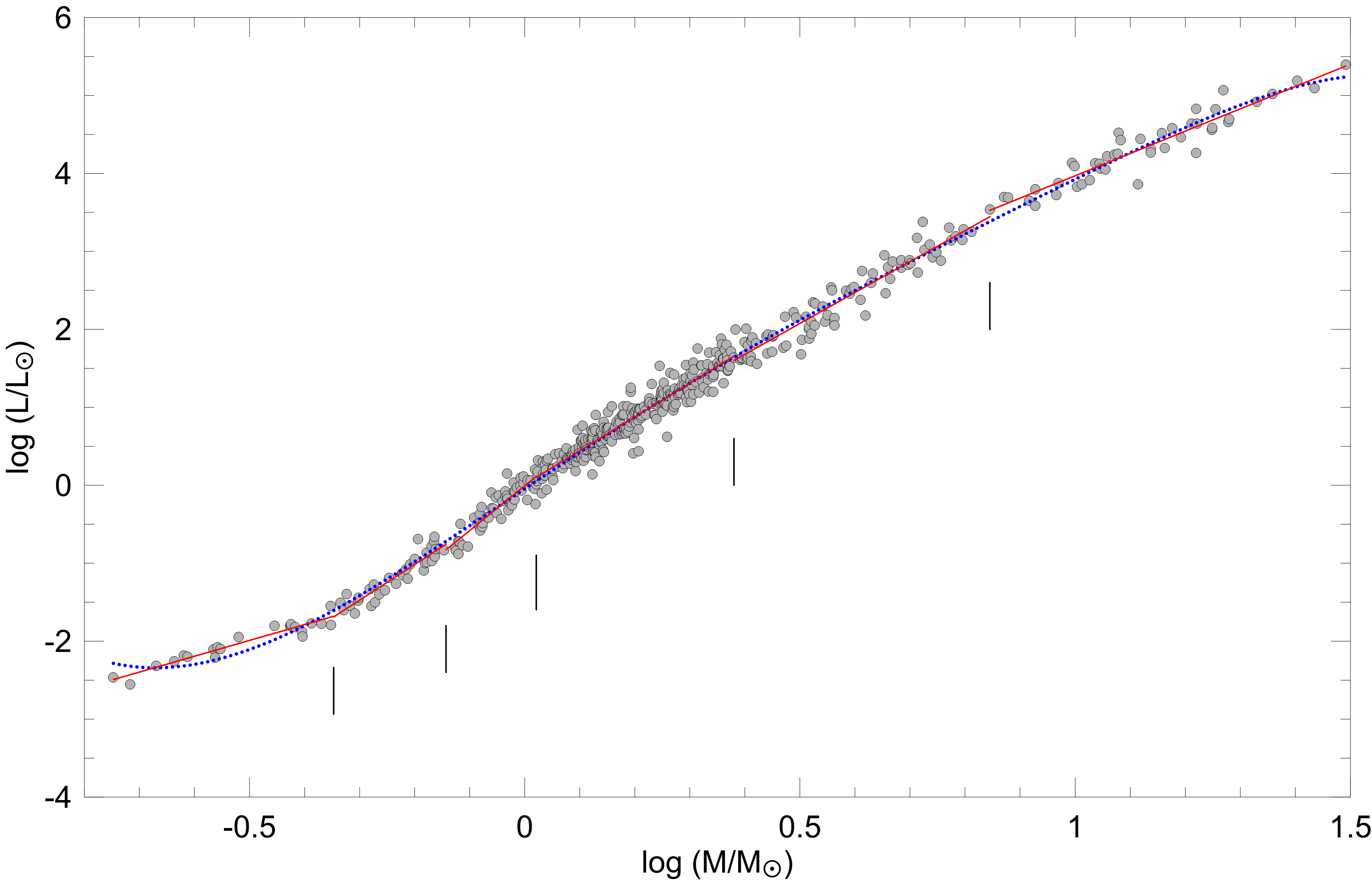}
\caption{The dotted (blue) line is a sixth-degree polynomial, solid (red) lines are classical MLRs, and the vertical lines are the break points separating mass domains where the linear lines were fitted \citep[credit to][]{Eker2018}.}
\label{fig:logL-logM}
\end {figure}

The distribution of the luminosities of 509 main-sequence stars on a $\log M-\log L$ diagram is shown in Figure~\ref{fig:logL-logM}, where the vertical lines mark the positions of the break points. Between the break points there is a linear MLR luminosities of that domain, thus there exists six-piece MLR to cover the full range $0.179 < M/M_{\odot}\leq 31$. A six-degree polynomial, shown by a blue dotted line to fit the full range, was found producing a best fit better than the higher and the lower degree polynomials to explain the full range data by a single function. It is obvious in Figure~\ref{fig:logL-logM} that the six-piece MLRs are even better at representing stellar luminosities than any polynomial of any degree. 

\begin{figure}
\centering
\includegraphics[scale=0.70, angle=0]{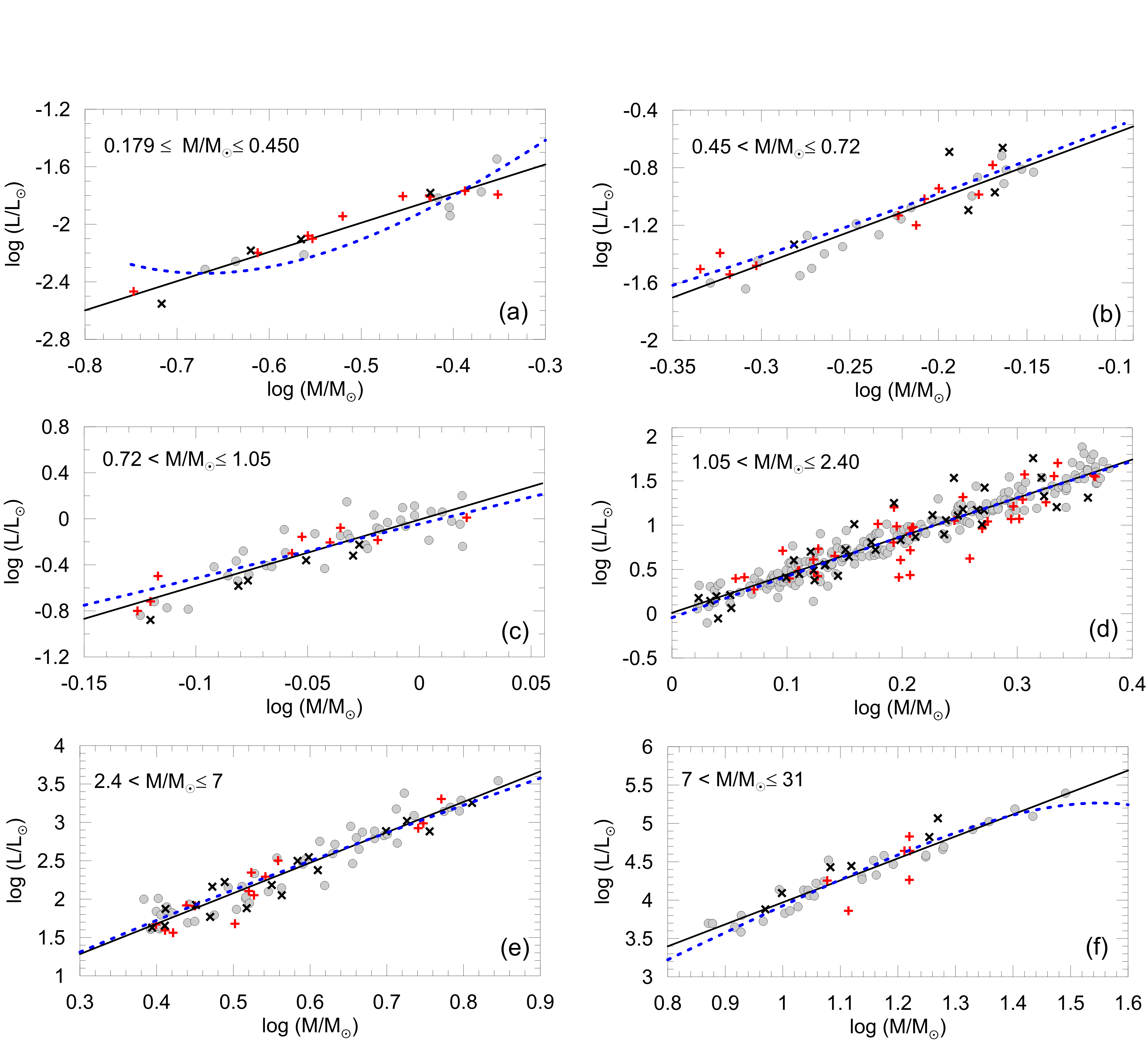}
\caption{Mass domains and classical MLRs representing stellar masses and luminosities in each domain. 6th degree polynomial representing all data (same as the dotted blue line in Figure \ref{fig:logL-logM}) is not as successful as linear MLRs in low-mass domains ($M<1.05$) and very high-mass domain ($M>7$). Data accuracy is: (o) very accurate $<3\%$, ($+$) accurate (3-6\%), and ($\times$) less accurate (6-15\%) \citep[credit to][]{Eker2018}.} 
\label{fig:logL-logM_range}
\end {figure}

The six-mass domains, which are named ultra-low mass ($0.179 < M/M_{\odot} \leq 0.45$), very-low mass ($0.45 < M/M_{\odot} \leq 0.72$), low mass ($0.72 < M/M_{\odot} < 1.05$), intermediate mass ($1.05 < M/M_{\odot} \leq 2.4$), high mass ($2.4 <M/M_{\odot} \leq 7$) and very-high mass ($7 <M/M_{\odot} \leq 31$), and their best fitting MLR functions are shown in Figure~\ref{fig:logL-logM_range}, where they were also compared to a single MLR function of sixth degree covering the whole mass range $0.179 < M/M_{\odot} \leq 31$. Analytical expressions, and statistical parameters $N$ (number of stars in the domain), $R^2$ (correlation coefficient), $\sigma$ (standard deviation), and $\alpha$ (the power of $M$) as the inclination of a linear MLR on a $\log M-\log L$ diagram is listed in Table~\ref{table:Massrange}.  

\begin{table}
\setlength{\tabcolsep}{5pt}
\centering
\small
\caption{~Classical MLRs for main-sequence stars at various mass domains \citep[credit to][]{Eker2018}.}
\begin{tabular}{lclcccc}
\hline
Domain            & $N$  & Mass Range & Classical MLR & $R^2$ & $\sigma$     & $\alpha$ \\
\hline
Ultra low-mass     &  22  & $0.179<M/M_{\odot}\leq0.45$  & $\log L=2.028(135)\times\log M-0.976(070)$ & 0.919 & 0.076 & 2.028 \\
Very low-mass      &  35  & $0.45< M/M_{\odot}\leq0.72$  & $\log L=4.572(319)\times\log M-0.102(076)$ & 0.857 & 0.109 & 4.572 \\
Low mass           &  53  & $0.72< M/M_{\odot}\leq1.05$  & $\log L=5.743(413)\times\log M-0.007(026)$ & 0.787 & 0.129 & 5.743 \\
Intermediate mass & 275  & $1.05< M/M_{\odot}\leq2.40$  & $\log L=4.329(087)\times\log M+0.010(019)$ & 0.901 & 0.140 & 4.329 \\
High mass         &  80  & $2.4 < M/M_{\odot}\leq 7$    & $\log L=3.967(143)\times\log M+0.093(083)$ & 0.907 & 0.165 & 3.967 \\
Very highmass    &  44  & $7< M/M_{\odot}\leq31$       & $\log L=2.865(155)\times\log M+1.105(176)$ & 0.888 & 0.152 & 2.865 \\
\hline
\end{tabular}%
\label{table:Massrange}%
\end{table}%

\subsection{Revisiting MRR}
Although it appears to be natural to sense a relation between $M$ and $R$ of main-sequence stars right after the discovery of MLR by \citet{Hertzsprung1923} and \citet{Russell1923}, the empirical MRR of main-sequence stars did not appear in the literature for another one and a half decades. \citet{Kuiper1938}, who was suspicious about MLR to be one of the fundamental relations, also plotted a $\log M-\log R$ diagram but had no discussion about it. Only after mid of the 20th century, the studies discussing empirical stellar mass-radius relation (MRR) begin to appear in the literature \citep{McCrea1950, Plaut1953, Huang1956, Lacy1977, Lacy1979, Kopal1978, Patterson1984, Gimenez1985, Harmanec1988, Demircan1991}. 

If $M$ and $L$ are related, why not $M$ and $R$ since $L$ is already known to be related to the square of $R$ according to the Stefan-Boltzmann law? The main obstacle seems to be the difficulty of accessing reliable stellar radii. \citet{McCrea1950} appears to be one of the first founders of the idea for two reasons: {\it i}) Using the light curve solutions of eclipsing binaries, not necessarily all being detached, that is, including W UMa binaries too, \citet{Plaut1953} has drawn a diagram in the manner proposed by \citet{McCrea1950}. {\it ii}) Readers were referred to \citet{McCrea1950} for theoretical interpretation of the relation in the form $\log (M/R)=a+b\log M$ suggested by \citet{Plaut1953}. Using the parameters of 30 systems with both components on the main sequence chosen among 130 eclipsing binaries of all kinds compiled by \citet{Plaut1953}, the coefficients of the relation were rectified as $a=-0.058\pm 0.026$, $b=0.335\pm 0.29$ with a dispersion of $\pm 0.11$ according to the least squares.  

The masses and radii of the eclipsing stars compiled by \citet{Plaut1953} had been studied by \citet{Huang1956}, who plotted them on the two separate spectral type-$\log R$ diagrams, on which both components smaller thus not touching, both components are touching, primary is touching but secondary is not, and the secondary is touching but the primary is not touching the inner contact surfaces were marked by special symbols for each. The two separate diagrams made \citet{Huang1956} believe that the sub-giant components in such systems could have evolved from early main-sequence stars. 

An empirical MRR in the form $\log R (M) = 8.495-0.2 M_{\rm Bol}(M)-2\log T_{\rm eff}(M)$ suggested by \citet{Hoxie1973} for the low mass stars ($M<1 M_{\odot}$) in the Solar neighbourhood. Comparing this MRR with low mass model calculations of stellar $R$, \citet{Hoxie1973} announced a discordance implying theoretical radii are 30\% or smaller than the observationally derived radii in the mass-radius plane for stars of mass less than $0.5 M_{\odot}$. Afterward, a method of estimating $R$ of nearby stars was declared by \citet{Lacy1977}. Being based on the Barnes-Evans relation and being free of assumptions of spectral types, luminosity class, effective temperature, or bolometric correction, this method is applied to nearby single and double stars with accurate parallaxes and $V-R$ photometry. The double stars (visual binaries) were useful to supply components $M$ and $R$, where $R$ needed to be compared with predicted $R$ by the new method with Barnes-Evans relation. Additional comparison was also possible to \citet{Lacy1977} using $R$ from non-contact eclipsing binaries with well-determined dimensions selected from \citet{Batten1967} and \citet{Koch1970}. Having a sufficient number of accurate masses and radii (both 5\% or better) from the visual systems and non-contact eclipsing binaries, \citet{Lacy1977} was also able to produce the most reliable $\log M-\log R$ diagram so far, where the empirical relation of \citet{Hoxie1973} was plotted with a special symbol different from the symbols of visual binaries, eclipsing systems undergone mass exchange and without any mass exchange.

Therefore, \citet{Lacy1977} suggested $\log R=0.640\log M+0.011$ for the region $0.12\leq \log M \leq 1.30$ and $\log R = 0.917 \log M-0.020$ for the region $-1.00\leq \log M\leq 1.30$ as two MRR functions to indicate zero age main sequence (ZAMS), where $R$ and $M$ are in Solar units. \citet{Lacy1977} interpreted the break point as a signal crossing over from the region of p-p chain to the C-N-O cycle. The terminal age main sequence (TAMS) line was deduced from the models of \citet{Iben1967} and \citet{Paczynski1970} marked on the $\log M-\log R$ as a dashed line, while ZAMS is shown by a continuous line with a break at $1.3\pm 1 M_{\odot}$. Theory and observation were found in good agreement for the stars $M \gtrsim 1M_{\odot}$, but the models of $M$ dwarfs having 25\% smaller radii than real stars.      
Another line of development was using the Catalogue of the Elements of Eclipsing Binaries \citep{Kopal1956}, where \citet{Kopal1959} obtained a good statistical definition of the MRR. He found a linear relation, in logarithmic scales, for the values of individual stars irrespectively being primary or secondary, but the slope is different for massive and less massive stars with a transition at $\sim 2 M_{\odot}$ \citep{Kopal1956, Kopal1978}. \citet{Popovici1974}, who compiled data from \citet{Kopal1956} and \textcolor{blue}{Svetchnikoff (1969)}\footnote{Catalog Orbitalnii Elementov, Mass i Svetimostii Tesnii Dvornik Svezd, Sverdlovsk}, were mainly interested in the radius-luminosity diagram. Five mean empirical mass-radius relations had been constructed by \citet{Habets1981}. The number of MRR is five because it is calibrated not only for main-sequence stars but also for luminosity class IV (subgiants), III (giants), $V_{\odot}$ (ZAMS) and $EV_{\odot}$ (EZAMS) using visual and eclipsing binaries which are at the same time double-lined spectroscopic systems collected by themselves. \citet{Patterson1984} was unsatisfied by the MRRs suggested before him, thus he proposed the empirical ZAMS mass-radius relation in the form $R=\alpha M^{\beta}$, where $M$ and $R$ in Solar units and $\alpha=1$, $\beta=0.88\pm 0.02$ for the region $0.1\leq M/M_{\odot}\leq 0.8$ and $\alpha= 0.98$, $\beta=1.00$ for the region $0.8\leq M/M_{\odot} \leq 1.4$ by a quotation ``poorly defined for $M\leq 0.4 M_{\odot}$'' for himself to apply it to cataclysmic variables. \citet{Gimenez1985} preferred using the Catalogue of Stellar Masses and Radii published by \citet{Popper1980} for studying the classical MRR on $\log M -\log R$ diagram. Using reliable $M$ and $R$ from OB and B6-M detached eclipsing and visual binaries and resolved spectroscopic binaries, \citet{Gimenez1985} determined $a=0.041\pm 0.011$ and $b= 0.749\pm 0.011$ for an MRR in the form $\log R = a + b \log M$ with a correlation coefficient 0.96. 

\begin{figure}
\centering
\includegraphics[scale=1.3, angle=0]{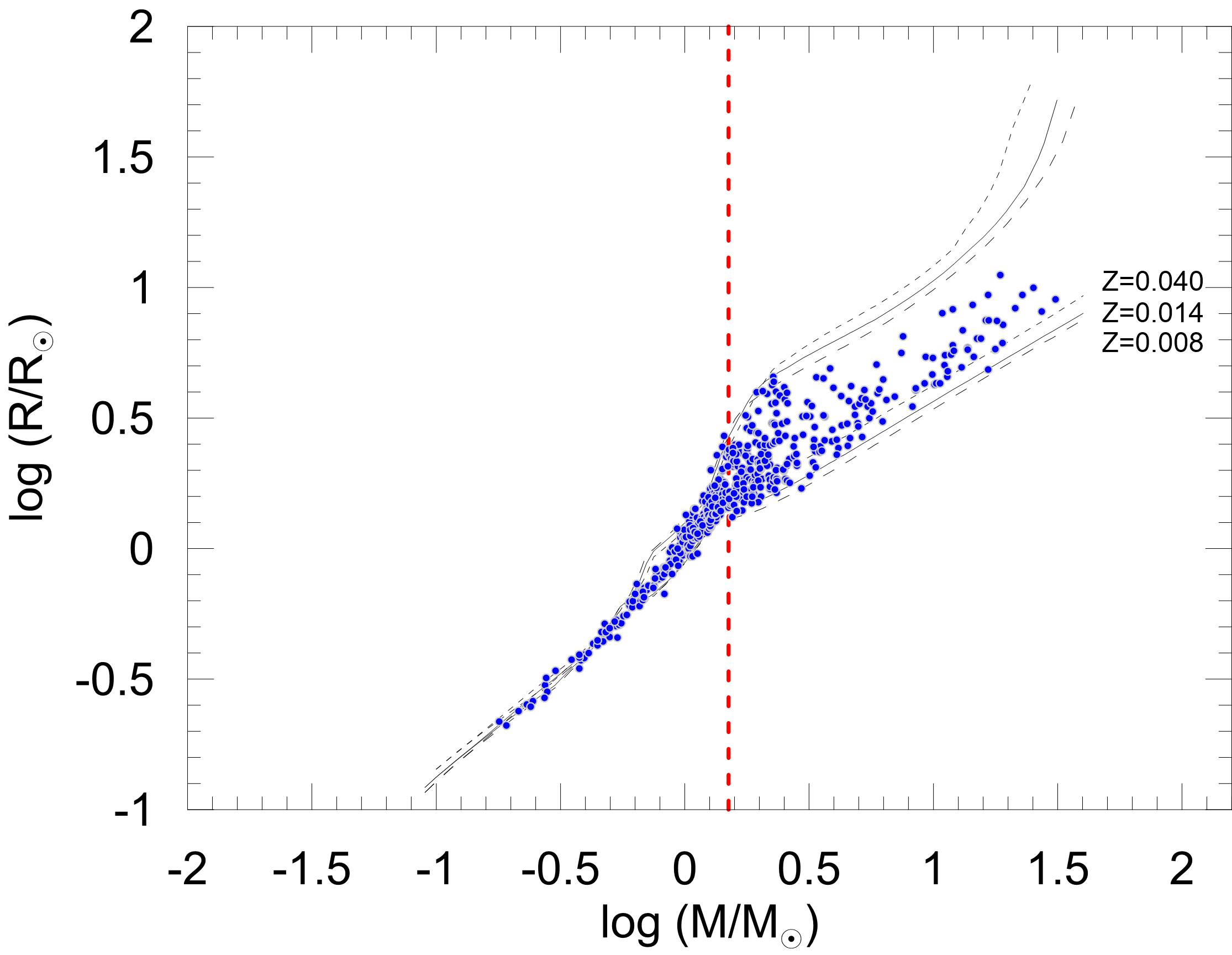}
\caption{Main sequence mass-radius diagram of DDEB stars. ZAMS and TAMS lines are from PARSEC models of \citet{Bressan2012} \citep[credit to][]{Eker2018}. The vertical line is a division at $M = 1.5 M_{\odot}$ ($\log M=0.176$).} 
\label{fig:logR-logM}
\end {figure}

It must be noted that \citet{Lacy1977, Lacy1979}, \citet{Gimenez1985} and \citet{Demircan1991} presented their $\log M- \log R$ diagrams without fitting a MRR function to data, but with a ZAMS line as the lower limit, and with a TAMS line as the upper limit of main-sequence stars which are estimated by the help of theoretical stellar structure and evolution models. The most recent example is by \citet{Eker2018} is shown in Figure~\ref{fig:logR-logM}, where it is clear that the statistical MRR does not appear as strong as classical MLR expressed on $\log M- \log L$ diagram (Figure~\ref{fig:logL-logM_range}). Looking at the difference of data distribution on both diagrams $\log M- \log L$ and $\log M- \log R$, therefore, it is not difficult to understand why $\log M- \log R$ diagram is presented without a fitting curve like Figure~\ref{fig:logL-logM_range}. 

Because of the very large scatter caused by the evolution of stars $M\gtrsim 1.3 M_{\odot}$ ($\log M > 0.15$), data of main-sequence stars on a $\log M- \log R$ diagram are not suitable to be expressed by a function or a curve. Obviously, plotting a ZAMS MLR on a $\log M -\log R$ diagram permits one to see the evolution of stellar radii for main-sequence stars $M\gtrsim 1.3 M_{\odot}$. 

Despite its hassle-free appearance, tight and narrow distribution of stellar $L$ on a $\log M- \log L$ diagram for the full mass range of main-sequence stars, different forms of MLR functions causing confusion were suggested, calibrated, and shown together with data. Looking at the already suggested main sequence MLR and MRR functions so far, one should see another noticeable difference. Being contrary to different forms of MLR suggested by different study groups, main sequence MRR in the form $\log R = a + b \log M$ happened to be common. The form $R\propto M^{\beta}$ suggest by \citet{Patterson1984}, \citet{Demircan1991} and \citet{Karetnikov1991}, appears different, but it is not. The same form of function is expressed once in logarithmic form while the other is a power law.

Despite such obvious differences (appearance on a diagram, formal expression), similarities between MLR and MRR studies also exist. \citet{Andersen1991}’s exclamation against classical MLR also affected MRR studies, thus modern MRR also appears to be deviated from the main path. Older MRR studies aimed to calibrate an MRR relation for two practical purposes: {\it i}) to get an answer to the question: how mean (or typical) $M$ and $R$ of main-sequence stars are related? {\it ii}) to estimate $R$ of a star for a given $M$, or vice versa. The former is useful for making models that require mean $M$ and $R$, the latter is useful for guessing the $R$ of a star from its $M$, or vice versa. Models using mean $M$ and $R$ as well as mean $L$, have larger application area not only in stellar astrophysics but also Galactic and extra-galactic studies.

The exclamation of \citet{Andersen1991}, unfortunately, caused the first aim to be neglected. As explained above, in the new trend of calibrating MLR, there is only one aim: to obtain the most accurate $M$ and $R$ together from the other observed parameters of the star. Thus, \citet{Malkov2007}, \citet{Torres2010}, \citet{Moya2018}, and \citet{Fernandes2021} have not only calibrated a single relation for predicting $R$ of a single star, but also at least one more relation useful to deliver $M$ of the same star. Thus, the new trend of MRR cannot be considered independent of the new trend of MLR appearing after \citet{Andersen1991}.         	

\subsection{Revisiting MTR}
If $M$ and $L$ are related, why not $M$ and $T_{\rm eff}$ since $L$ is known to equal surface area times surface flux, where the flux is proportional to the fourth power of the effective temperature. A relation even stronger than mass-radius is expected because the power of $T_{\rm eff}$ is two times bigger than the power of $R$ in the Stefan-Boltzmann law. Its calibration and usage also appear easier than the calibration and usage of MRR relation. Despite all, if the spectral type-effective temperature (horizontal axis of H-R diagram) relation is not counted, it took almost six decades for an MTR to be seen in the literature. 

First empirical mass-$T_{\rm eff}$ relation is studied by \citet{Habets1981} and shown on a $\log M - \log T_{\rm eff}$ diagram for main sequence-stars after studying newly calibrated mass-spectral type and spectral type-$T_{\rm eff}$ relations from the main-sequence components of eclipsing binaries. Empirical relation was shown together with two theoretical mass-$T_{\rm eff}$ relations of \citet[][for a metallicity $Z=0.02$ with $Y=0.49$ and 0.25]{Stothers1974}  and the one by \citet[][for $Z=0.02$ and $Y=0.25$]{Demarque1975}. The form of the relation was not given. \citet{Karetnikov1991} later determined coefficients of mass-$T_{\rm eff}$ relation in the form $\log T_{\rm eff} = a + b\log M$ for six different kinds of eclipsing systems using absolute parameters of 303 eclipsing binaries of different types with varying and constant orbital periods without showing them on diagrams. The relations were determined for primaries and secondaries separately, thus coefficients of 24 relations were determined and listed without showing them on $\log M - \log T_{\rm eff}$ diagrams. 
 
Apparently, MTR relation had also been affected by \citet{Andersen1991}'s objection to MLR because \citet{Malkov2007} announced a direct $[\log T_{\rm eff} - (\log M)]$ and an inverse $[\log M - (\log T_{\rm eff)}]$ mass-effective temperature relations together with his $M$ and $R$ predicting relations and their inverse functions. Like $\log M- \log L$ diagram without a fitting curve, his $M$ versus $\log T_{\rm eff}$ diagram too presented without a fitted curve. Being not interested in calibrating an MTR, \citet{Moya2018}, on the other hand, used $T_{\rm eff}$ as one of the free parameters for predicting $M$ and $R$ of single stars. \citet{Moya2018}, calibrated 38 relations: 18 for $M$ and 20 for $R$. This is very much similar to \citet{Torres2010} who suggested only two relations using $T_{eff}$, $\log g$ and [Fe/H] as free parameters, one for obtaining $M$ one for obtaining $R$, to replace classical MLR claimed inadequate to provide stellar $M$ by two reasons: {\it i}) It is a mean relation, thus $M$ from $L$ is very inaccurate. {\it ii}) Scatter on $\log M- \log L$ diagram is not only due to random observational errors of $M$ and $L$ but also due to stellar age and chemical composition differences. As if, a reliable (or true) MLR function must contain all of the parameters which introduce scattering. Such a MLR function, however, cannot be drawn on a $\log M- \log L$ diagram.

On the other hand, following the old tradition, that is, looking for mean relations for MRR and MTR, \citet{Eker2015} have compared $L$, $R$, and $T_{\rm eff}$ distributions on $\log M- \log L$, $\log M- \log R$ and $\log M - \log T_{\rm eff}$ diagrams. The first comparison between $\log M - \log L$, $\log M -\log R$ has been commented as ``the appearance of data on the $\log M- \log R$ diagram is very different than the appearance on the $\log M- \log L$ diagram (compare Figures~\ref{fig:logL-logM} and \ref{fig:logR-logM}), which rather looks like a band of data expressible by a function; however, with a very narrow distribution of radii for masses $M < 1 M_{\odot}$ and a broad band of radii for stars with $M > 1 M_{\odot}$, a single function to express a MRR would be odd and meaningless''. Then, comparison between $\log M- \log R$, $\log M- \log T_{\rm eff}$ were commented as ``the temperature evolution within the main-sequence band is not that obvious on the $M- T_{\rm eff}$ diagram. At first look, it resembles the MLR''. It is indeed not like MRR, where the main-sequence evolution of $R$ is obvious for $M > 1 M_{\odot}$ (compare Figures~\ref{fig:logL-logM}, ~\ref{fig:logR-logM}, and  \ref{fig:logT-logM}). 

Despite the distribution of data on $\log M - \log T_{\rm eff}$ diagram (Figure~\ref{fig:logT-logM}) resembles the distribution of data on $\log M- \log L$ diagram, \citet{Eker2015} preferred not to calibrate a mean MTR since it would have been odd or inappropriate that time to oppose literature where MLR and MRR are many, but MTR is almost absent. Later, \citet{Eker2018} noted “Stefan–Boltzmann law clearly indicates stellar luminosities are related to stellar radii and effective temperatures. Having empirically determined the MLR and MRR available, one is not free to determine another independent mass–effective temperature relation (MTR)”. The three independently calibrated MLR, MRR, MTR functions are not guaranteed to give consistent $L$, $R$ and $T_{\rm eff}$ for a given $M$. The solution to the problem is in the next subsection. 

\subsection{Revisiting interrelated MLR, MRR, MTR }

It has been noticed that the distribution of $R$ on the $\log M - \log R$ diagram (Figure~\ref{fig:logR-logM}) on the left of the vertical line ($M\leq1.5 M_{\odot}$) is smooth and tight, that is, it is expressible by a simple function which is to be called MRR. But the high mass region $M>1.5 M_{\odot}$ due to faster evolution, $R$ values are scattered very much, and thus, its band-like appearance does not seem expressible by a curve of a function.

\begin{figure}
\centering
\includegraphics[scale=1.00, angle=0]{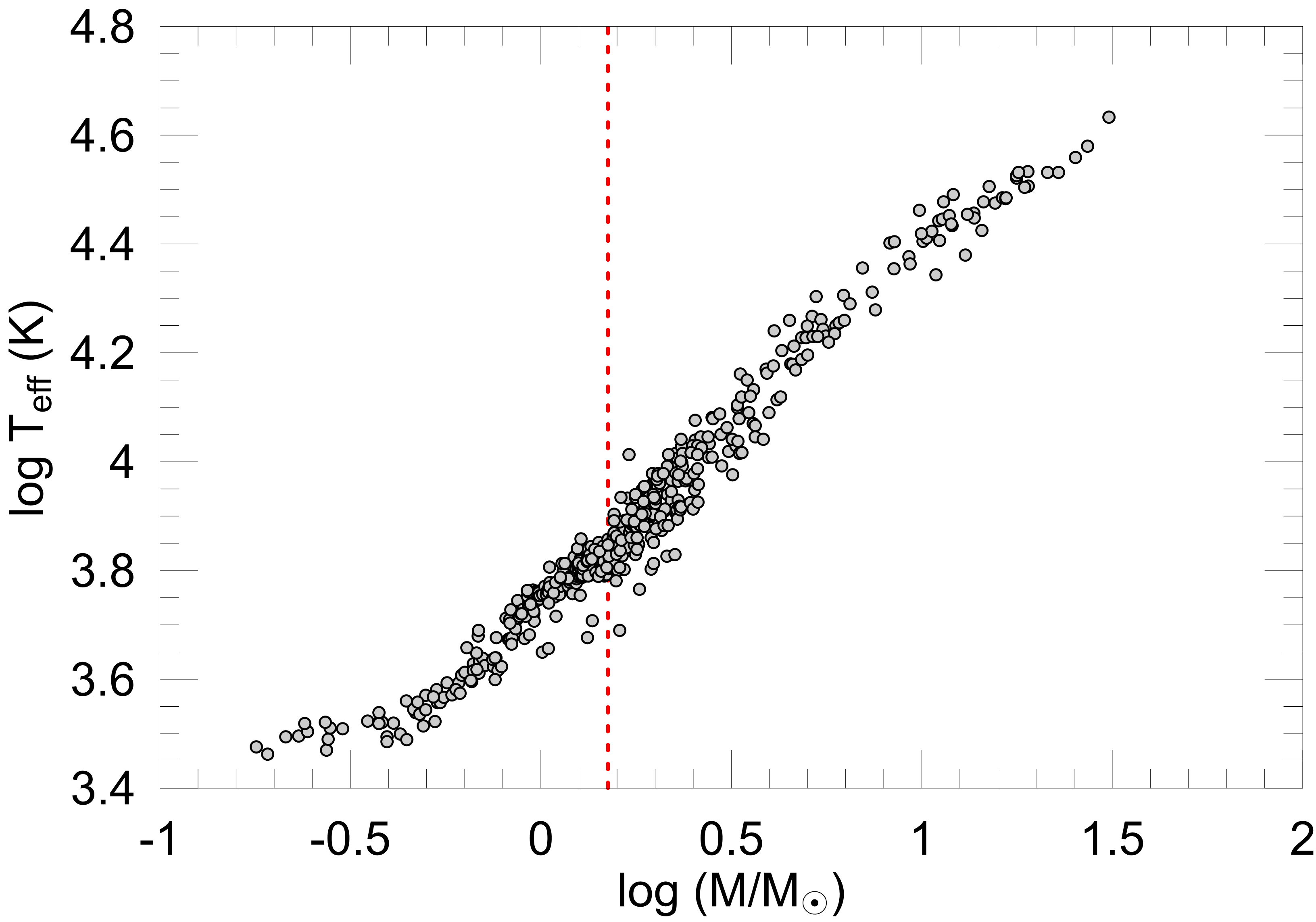}
\caption{~Main sequence mass-effective temperature diagram of DDEB stars \citep[credit to][]{Eker2018}. The vertical line is a division at $M = 1.5 M_{\odot}$ ($\log M=0.176$).} 
\label{fig:logT-logM}
\end {figure}

Despite its overall appearance resembling an MLR, the $T_{\rm eff}$ distribution on the $\log M - \log T_{\rm eff}$ diagram shows almost opposite characteristics for one to choose the same mass domain to calibrate a MTR. One can easily notice that the low-mass region $M\leq1.5 M_{\odot}$ (left of the vertical line) in Figure~\ref{fig:logT-logM} has a tight but wavy distribution, which is rather not possible to fit a simple function. On the contrary, one can easily spot the domain of high-mass stars $M>1.5 M_{\odot}$ (right of the vertical line) in Figure~\ref{fig:logT-logM} with a sufficiently tight but smoothly varying distribution, which is easily expressible by a function to be called MTR.

Having six linear MLRs already calibrated (Table~\ref{table:Massrange}, Figures~\ref{fig:logL-logM} and \ref{fig:logL-logM_range}) for the main-sequence stars in the full range of observed stellar masses, probable inconsistencies among the mean $L$, $R$, and $T_{\rm eff}$ values for a given $M$ will be eliminated if the MRR was calibrated for the low-mass region $M\leq 1.5 M_{\odot}$ only and the MTR was calibrated for the high-mass region $M>1.5 M_{\odot}$ only. This is because the vertical line is a dimensionless border between the low and the high mass regions in both $R$ and $T_{\rm eff}$ distributions as shown in Figures~\ref{fig:logR-logM} and \ref{fig:logT-logM}. Otherwise, with independently calibrated MLR, MRR, and MTR functions, one will have a mean luminosity ($\langle L\rangle$), a mean radius ($\langle R\rangle$), and a mean effective temperature ($\langle T_{\rm eff}\rangle$) for a given $M$. No one would know which ones of the three mean values are wrong because the mean $\langle L\rangle$ will not be equal to a mean surface area $\langle 4\pi R^2\rangle = 4\pi (\langle R^2\rangle)$ multiplied by a mean surface flux $\langle \sigma T^4\rangle = \sigma(\langle T_{\rm eff}^4\rangle)$ for a typical main-sequence star of given $M$. Choosing the most eligible regions on $\log M - \log R$ and $\log M - \log T_{\rm eff}$ diagrams as compensating mass domains for covering the full mass range, not only removes a probable inconsistency but also guarantees the most trustable MLR and MRR for the low-mass stars ($M\leq 1.5 M_{\odot}$) and the most trustable MLR and MTR for the high-mass stars ($M>1.5 M_{\odot}$). Then, one could calculate consistent $\langle L\rangle$, $\langle R\rangle$ and $\langle T_{\rm eff}\rangle$ for the full range of main-sequence stars, which happens in two steps for both of the mass domains. For the low-mass stars: {\it i}) Use MLR and MRR to calculate $\langle L\rangle$ and $\langle R\rangle$ for a given mass, {\it ii}) use Stefan-Boltzmann law to calculate $\langle T_{\rm eff}\rangle$ for the same mass from its already computed $\langle L\rangle$ and $\langle R\rangle$. For the high-mass stars: {\it i}) Use MLR and MTR to calculate $\langle L\rangle$ and $\langle T_{\rm eff}\rangle$ for a given mass, {\it ii}) use Stefan-Boltzmann law to calculate $\langle R\rangle$ for the same mass from its already computed $\langle L\rangle$ and $\langle T_{\rm eff}\rangle$. 

Utilizing the least squares, \citet{Eker2018} determined an empirical MRR directly from $M$ and $R$ of 233 main-sequence stars for low-mass stars within range $0.179\leq M/M_{\odot}\leq 1.5$, and an empirical MTR from $\log M$ and $\log T_{\rm eff}$ of 276 main-sequence stars for high-mass stars within range $1.5< M/M_{\odot}\leq 31$. Because the table giving the empirical MRR and MTR by \citet{Eker2018} has misprints, the open forms of the functions, correlation coefficients ($R^2$), and standard deviations ($\sigma$) from \citet{Eker2021c} are given here. The empirical MRR in the form of quadratic equation, where $R$ and $M$ are Solar units, is  

\begin{equation}
    R = 0.438(0.098)\times M^2 +0.479(0.180)\times M + 0.137(0.075),
\end{equation}
has $R^2=0.867$ and $\sigma = 0.176$. The empirical MRR in the form of a quadratic equation, where $T_{\rm eff}$ and $M$ are Kelvin and Solar units, is 

\begin{equation}
    \log T_{\rm eff} = -0.170(0.026)\times (\log M)^2 +0.888(0.037)\times \log M + 3.671(0.010), 
\end{equation}
has $R^2 = 0.961$ and $\sigma = 0.042$.

\begin{figure}
\centering
\includegraphics[scale=0.95, angle=0]{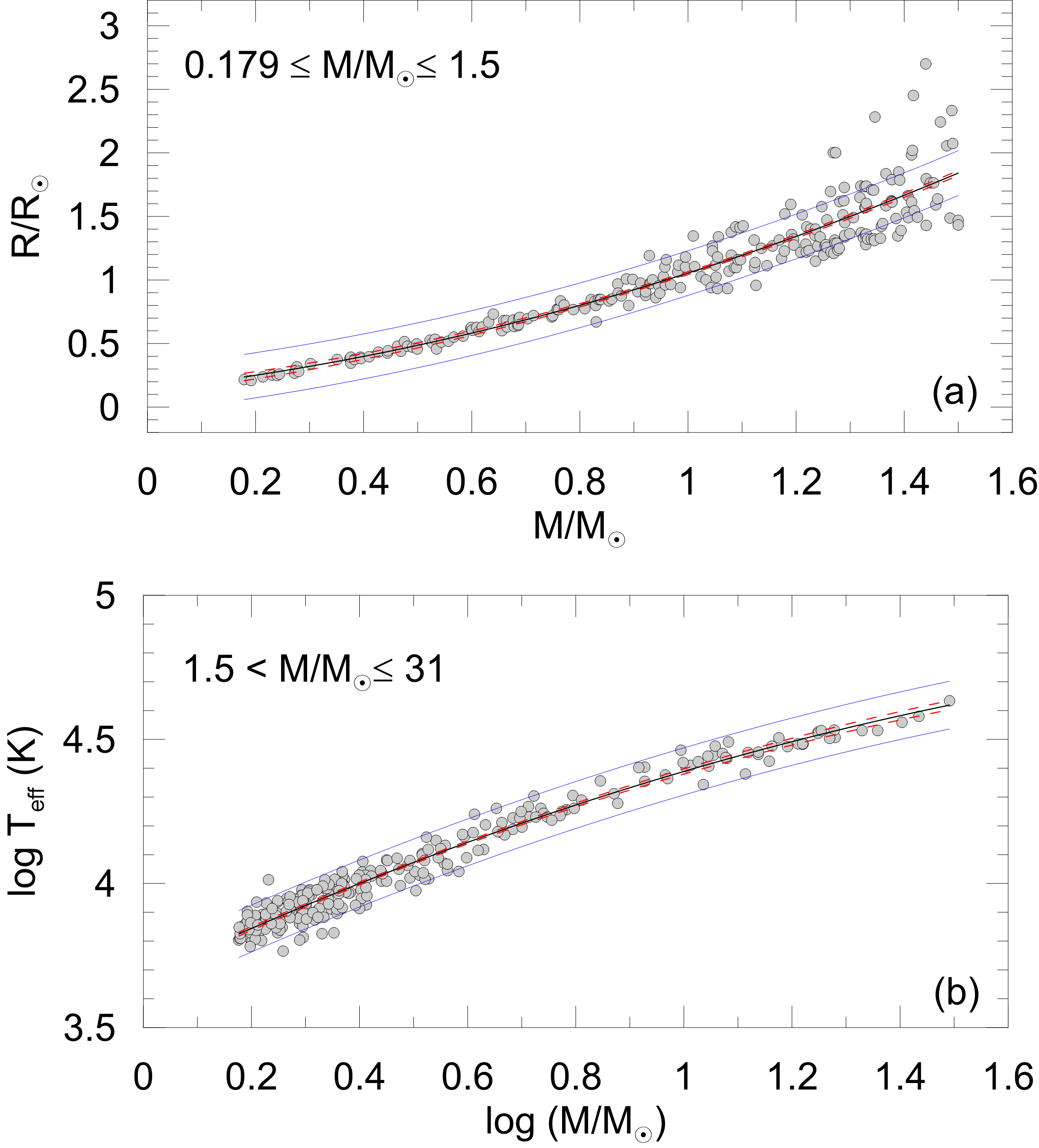}
\caption{Empirical MRR and MTR calibrated from 509 main-sequence stars for the mass ranges $0.179\leq M/M_{\odot}\leq 1.5$ and $1.5\leq M/M_{\odot}\leq 31$ of DDEB stars. Note MTR is in logarithmic scale, but MRR is not \citep[credit to][]{Eker2018}.} 
\label{fig:R-T-M}
\end {figure}

The empirical MRR and MTR by \citet{Eker2018} are shown together with data in Figure~\ref{fig:R-T-M}. The division between the low and high mass stars at $1.5 M_{\odot}$ is a nice coincidence that plots of MRR and MTR appear with the same numbers in the horizontal axis, thus one must be careful MTR is shown in logarithmic scale while MRR is not. 

\begin{figure}
\centering
\includegraphics[scale=0.9, angle=0]{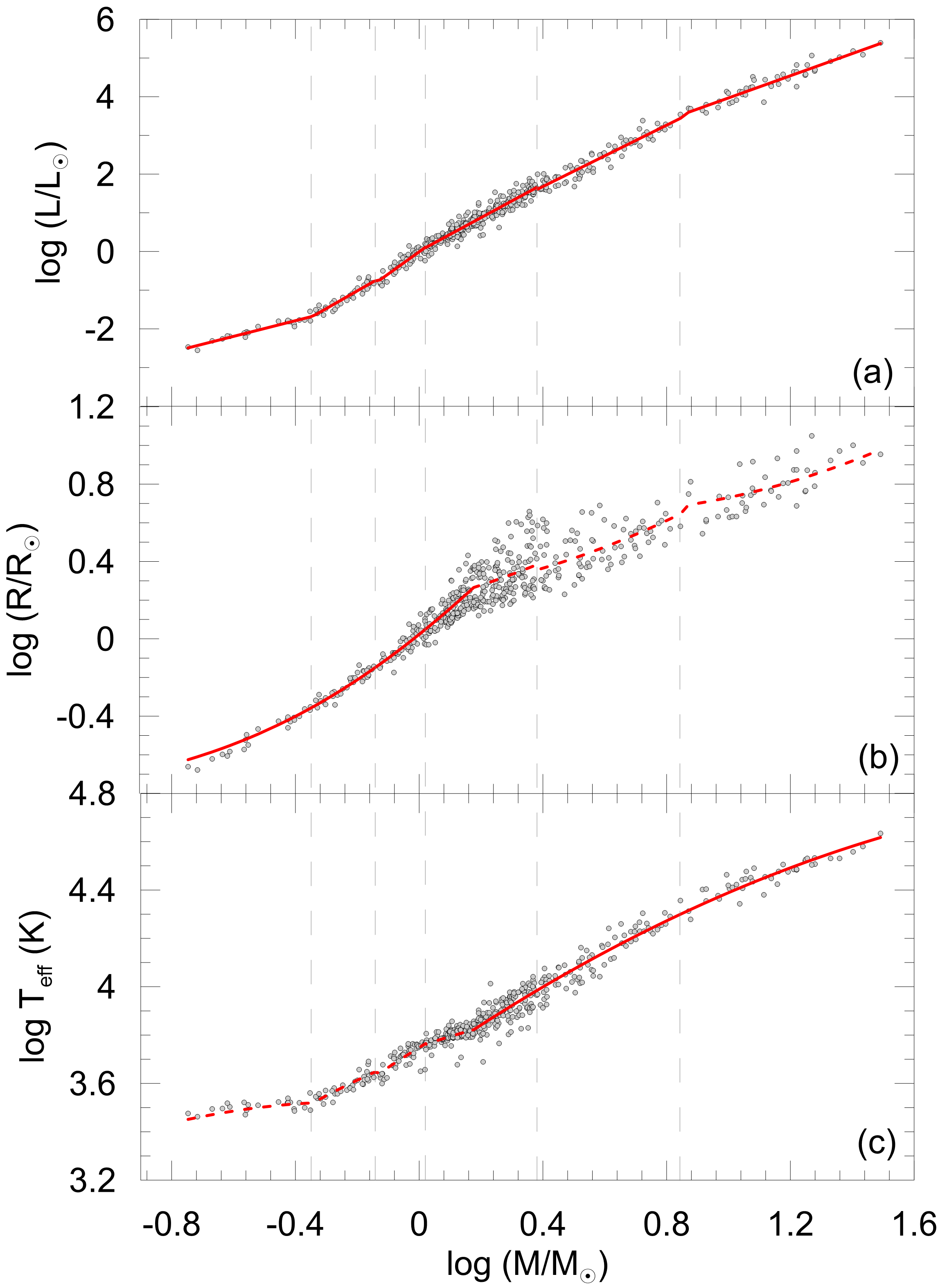}
\caption{The mean values ($\langle L\rangle$, $\langle R \rangle$, $\langle T_{\rm eff}\rangle$) computed from empirical MLT, MRR and MTR are compared to $L$, $R$ and $T_{\rm eff}$ of 509 main-sequence stars. The mean values directly from MLR, MRR or MLR, MTR are solid, The mean values from the Stefan-Boltzmann law are dotted. Dashed vertical lines are the break points of MLR \citep[credit to][] {Eker2018}.}
\label{fig:L-R-T-M}
\end {figure}

The mean values ($\langle L\rangle$, $\langle R\rangle$, $\langle T_{\rm eff}\rangle$) computed from the empirical MLR, MRR and MTR of \citet{Eker2018} are compared to $L$, $R$ and $T_{\rm eff}$ of 509 main-sequence stars chosen from 318 DDEB and one detached eclipsing triple are shown in Figure~\ref{fig:L-R-T-M}. The solid lines mark the mean values directly from MLR, MRR or MLR, MTR, while the dotted lines mark the mean values $\langle T_{\rm eff}\rangle$ calculated from $\langle L\rangle$, $\langle R\rangle$ for the low-mass stars and $\langle R\rangle$ calculated from $\langle L\rangle$ and $\langle T_{\rm eff}\rangle$ for the high-mass stars.

Calculating the mean values ($\langle L\rangle$, $\langle R\rangle$, $\langle T_{\rm eff}\rangle$) in two steps as described above has an advantage further checking them up and confirming whether the mean values of the first step is consistent or not. Note that, only if $\langle L\rangle$ and $\langle R\rangle$ values of the low mass stars are consistent, then in the second step consistent $\langle T_{\rm eff}\rangle$ values could be produced from $\langle L\rangle$ and $\langle R\rangle$; similarly, only if $\langle L\rangle$ and $\langle T_{\rm eff}\rangle$ values of the high-mass stars are consistent, then in the second step, consistent $\langle R\rangle$ values could be produced according to the Stefan-Boltzmann law. The three break points and varying inclinations of MLR before, between and after the break points in the mass range $M\leq 1.5 M_{\odot}$, apparently, causing the wavy look of the MTR of low-mass stars. What simple function would have produced the current successful appearance of the fit displayed in Figure~\ref{fig:L-R-T-M}c by the dotted line? Similar influence of the break points and effect of varying inclinations of MLR before, between and after the break points for the high mass stars $M>1.5 M_{\odot}$ are there in the middle panel (Figure~\ref{fig:L-R-T-M}b) but appears to be lost within the scatter caused by the faster evolution of more massive stars. Therefore, it could be said that it is a bull’s eye to choose the mass region $M\leq 1.5 M_{\odot}$ for devising MRR and to choose the mass region $M > 1.5 M_{\odot}$ for devising MTR. 

The two-step procedure in determining the mean values of $\langle L\rangle$, $\langle R\rangle$, $\langle T_{\rm eff}\rangle$ for a given $M$ may appear problematic to introduce extra errors in the error propagation. Users should not be deceived from this illusion. First of all, this is because, consistent mean values of higher uncertainty is better than inconsistent more accurate ones. To avoid inconsistency, one may use two-step procedure as described above or will use MLR and MRR only or MLR and MTR only for the full mass range of main-sequence stars. Thus, in either case, the two-step procedure is inevitable for consistent results. For the high mass stars $M > 1.5 M_{\odot}$ the scatter on $R$ is too big thus authors usually prefer not to determine MRR. Then for the former case, MRR and MTR would be missing for the high mass stars while interrelated MLR, MRR and MTR would be possible only for low mass stars. For the low mass stars $M < 1.5 M_{\odot}$, on the other hand, it is not possible to find a simple function to fit in determining MTR. Then, interrelated MLR, MRR and MTR becomes possible only for the high mass stars since MRR and MTR would be missing for the low mass stars $M < 1.5 M_{\odot}$. if one prefers to fit a MRR together with MLR in the full mass region, then he/she will end up having very unreliable $\langle R\rangle$ for the stars $M > 1.5 M_{\odot}$. Error propagation, then, would produce similarly unreliable, even worse $\langle T_{\rm eff}\rangle$ since independent MTR function is not there. if one prefers to calibrate MLR and MTR in the full range of masses, similar unreliable or erroneous $\langle T_{\rm eff}\rangle$, $\langle R\rangle$ values emerge for the low mass stars $M < 1.5 M_{\odot}$. Therefore, the two-step procedure as described in the paragraph before this one in determining the mean values of $\langle L\rangle$, $\langle R\rangle$, $\langle T_{\rm eff}\rangle$ for a given $M$ always better and more reliable than the independently determined MLR, MRR and MTR for sure.

\section{Conclusions}

Fundamental relations are handy tools to explain simple natural phenomena or starting points for one to understand more complex happenings. The main difference between statistical and fundamental relations is that a statistical relation is valid under certain conditions implied by the data from which it was formulated while a fundamental relation is rather independent of data, that is, data is not there to constrain it but only to confirm or to falsify it. All empirical relations derived from observational or experimental data are statistical in spirit. This does not, however, mean a fundamental relation can not be derived from observational and/or experimental data. A statistical study could reproduce and even may find a fundamental relation.

Physical laws are indifferentiable from fundamental relations. Their scope is wider; for example, the Stefan-Boltzmann law ($L=4\pi R^2\sigma T_{\rm eff}^4$) is valid for all stars radiating thermally, also for a hypothetical star representing $\langle L\rangle$, $\langle R\rangle$, and $\langle T_{\rm eff}\rangle$ implied by main sequence MLR, MRR, and MTR. On the contrary, MLR, MRR, and MTR are valid to give mean values for a main-sequence star of a given $M$ within a valid mass range, which was suggested after their calibrations. As they are not valid for non-main sequence stars, they are also not valid to estimate $L$, $R$, or $T_{\rm eff}$ of a main-sequence star with a known $M$ only. This is because, nowadays many stellar structure and evolution models exist for one to look for $L$, $R$, and $T_{\rm eff}$ of a star according to its $M$, chemical composition ($X, Y, Z$) and age.

Because of their simplicity, some statistical relations could be mistaken or misused as a fundamental relation. A good example of this is the classical MLR. This is because, right after its discovery by \citet{Hertzsprung1923} and \citet{Russell1923}, for a while, there were no other methods to obtain masses of single stars but estimate it from $L$ or absolute bolometric magnitude ($M_{\rm Bol}$) by using a MLR in the forms of $L\propto M^{\alpha}$ or $M_{\rm Bol}$ - mass diagrams. 
 
Having no alternative is another property of fundamental relations. Indeed, the Stefan-Boltzmann law or the Planck law (spectral energy distribution of a black body with specific temperature) do not have alternatives. If there is no alternative to estimate $M$ of a single star, but using a classical MLR, convinced early astronomers that it could be used as a fundamental relation. Today, there are alternatives to getting $M$ and $R$ of single stars, thus there is no excuse to assume any form of MLR as a fundamental relation. One might still want to use MLR, MRR, and MTR, as a fundamental law, that is, he/she is more interested in knowing $M$ of star from its $L$. This is permissible with a larger uncertainty covering with of the main-sequence at the value of $M$ on a $\log M - \log L$ diagram, simply $\pm \sigma$, which is given in Table~\ref{table:Massrange}.    

On the other hand, statistical relations could be devised to serve a very specific purpose; for example, to obtain $M$ and $R$ of a single star from its other observables. In this respect, a statistical relation may appear to be operating like a fundamental relation, but this is an illusion. $M$ and $R$ predicting relations are still statistical relations if their validity depends on the data from which they were calibrated. Also, because there are many alternatives to provide $R$ or $M$ of a star from its $T_{\rm eff}$, $L$, $g$, $\rho$, and [Fe/H] as expressed by \citet{Moya2018} and as summarized by \citet{Serenelli2021}. \citet{Malkov2007}, \citet{Torres2010}, and \citet{Fernandes2021} have other alternatives. 

Naming a newly devised or renaming a re-calibrated empirical relation is very important; as important as classifying it as another statistical relation or one of the fundamental laws. Names must be unique to avoid confusion and guide users towards the original purpose of the relation. Repeating the same improper and non-unique names, should not be advocated by saying ``Let us keep the same name in the past''. This will be nothing but insisting on the same error. There are many miscalling or non-unique naming examples in the past, which are identified with possible correct names in the previous section. Especially after \citet{Andersen1991}'s exclamation that some $R$ and $M$ predicting relations were non-uniquely and incorrectly called MLR. Let us hope, \citet{Chevalier2023}, and \citet{Malkov2022} would be the last examples who erroneously named their mass-$M_{\rm G}$ diagrams, where $M_{\rm G}$ is absolute brightness in {\it Gaia} $G$ band, ``mass-luminosity'' diagram, and their mass – $M_{\rm G}$ relations ``mass-luminosity'' relation. Possible unique names for them are the ``mass-absolute brightness'' diagram or the ``mass-absolute brightness'' relation at {\it Gaia} $G$ band.            

The empirical relations for predicting $M$ and $R$ of single stars developed as alternatives to classical MLR and MRR after the split, initiated after \citet{Andersen1991}'s exclamation, occurred on the classical path aiming to get mean values $\langle L\rangle$, $\langle M\rangle$, and $\langle R\rangle$, and individual $L$, $R$, and $M$ of single stars. In this respect, it is not possible for one to say, ``the empirical relations giving $M$ and $R$ of single stars are more valuable than the classical MLR and MRR''. Claiming the opposite is also not correct. Both schools of thought have useful applications, valuable with respect to their own aims. Empirical $M$ and $R$ predicting relations had found a fruitful application to explore exoplanets hosting single stars. Estimating $M$ and $R$ of planet-hosting stars are the starting point for exploring hosted exoplanets \citep{Stassun2017, Stassun2018}. Classical MLR, MRR, and MTR, on the other hand, are practical for constructing astrophysical models that need mean values, not only beneficial to stellar astrophysics but also Galactic, extragalactic search, and even cosmological models.

Further improvements in the predicting accuracy of single stars parameters depend mainly on the quantity and the quality of the radial velocity and light curve solutions of DDEB. Advances on already established relations or new forms are possible due to increasing demand for exoplanet investigations. Developments on the classical MLR, MRR, and MRR, on the other hand, are encouraged to include the metallicity effect, as well as to extend them further towards high and low mass limits, not only for the main sequence but also for other luminosity classes.

\section*{Acknowledgements}
Authors thank the two anonymous referees who both gave valuable comments in improving the quality of discussions in the manuscript. 



\bibliographystyle{mnras}
\bibliography{refs}

\bsp	
\label{lastpage}
\end{document}